\documentclass[english,usenatbib]{mn2e}

\usepackage{url}
\usepackage[a4paper,margin=0.8in]{geometry}
\usepackage{aas_macros}
\usepackage[english]{babel}
\usepackage[utf8x]{inputenc}
\usepackage{todonotes}
\usepackage{algorithmic}
\usepackage{algorithm}
\usepackage{units}
\usepackage{amsmath}
\usepackage{graphicx}
\def\pvm#1{[PM: {\it #1}] }
\def\pvm2#1{}

\title{FastPM: a new scheme for fast simulations of dark matter and halos}

\author[Yu Feng, Man-Yat Chu, Uro\v s Seljak, Patrick McDonald]{
Yu Feng$^1$\thanks{mailto:yfeng1@berkeley.edu}, Man-Yat Chu$^1$, Uro\v s Seljak$^{1,2}$, Patrick McDonald$^2$
\\
$^{1}$Berkeley Center for Cosmological Physics, Department of Physics, University of California Berkeley, Berkeley CA, 94720\\
$^{2}$ Lawrence Berkeley National Laboratory, Berkeley, CA, 94705\\
}

\begin{document}
\maketitle

\begin{abstract}
We introduce FastPM, a highly-scalable approximated particle mesh N-body solver, which implements the particle mesh (PM) scheme enforcing correct linear displacement (1LPT) evolution via modified kick and drift factors. Employing a 2-dimensional domain decomposing scheme, FastPM scales extremely well with a very large number of CPUs. In contrast to COmoving-LAgrangian (COLA) approach, we do not require to split the force or track separately the 2LPT solution, reducing the code complexity and memory requirements. We compare FastPM with different number of steps ($N_s$) and force resolution factor ($B$) against 3 benchmarks: halo mass function from Friends of Friends halo finder, halo and dark matter power spectrum, and cross correlation coefficient (or stochasticity), relative to a high resolution TreePM simulation. We show that the modified time stepping scheme reduces the halo stochasticity when compared to COLA with the same number of steps and force resolution. While increasing $N_s$ and $B$ improves the transfer function and cross correlation coefficient, for many applications FastPM achieves sufficient accuracy at low $N_s$ and $B$. For example, $N_s=10$ and $B=2$ simulation provides a substantial saving (a factor of 10) of computing time relative to $N_s=40$, $B=3$ simulation, yet the halo benchmarks are very similar at $z=0$. We find that for abundance matched halos the stochasticity remains low even for $N_s=5$. FastPM compares well against less expensive schemes, being only 7 (4) times more expensive than 2LPT initial condition generator for $N_s=10$ ($N_s=5$). Some of the applications where FastPM can be useful are generating a large number of mocks, producing non-linear statistics where one varies a large number of nuisance or cosmological parameters, or serving as part of an initial conditions solver. 
\end{abstract}

\section{Introduction}

Extracting full information from observations of the 
large scale structure (LSS) of the universe, in weak lensing, galaxies, and other tracers, requires accurate predictions, which are only possible using simulations. These simulations can be used to create mock galaxy or weak lensing catalogs for co-variance estimation \citep{nifity2015}, to vary the predictions as a function of cosmological or nuisance parameters \citep[for example, galaxy formation parameters in halo occupation models as by][]{2002PhR...372....1C}, or even as a tool to generate initial conditions \citep{wang2013,2013MNRAS.432..894J,2013MNRAS.429L..84K}.
A full N-body simulation is too expensive both in terms of CPU-hours and wall-clock times. For this reason approximated N-body solvers that produces a reasonably accurate dark matter density field provides a practical alternative in critical applications. They range from simple Lagrangian perturbation theory field realizations \citep{patchy2014,pinocchio2013}, to N-body simulations optimized for specific applications. 

A large class of codes that employ the latter is the particle mesh (PM) family \citep[e.g.][]{pmfast2005,carlson09,coyote2010,pmlya2010,cola13,qpm14}.  The idea is to ignore the force calculation on small scales by skipping the Tree (PP) force part of a full Tree-PM (${\rm P^3M}$) code. Vanilla PM has been criticized for failing to reproduce the linear theory growth at large scale as the number of time steps is reduced \citep[see e.g.][]{2016MNRAS.459.2327I}. Recently, \cite{cola13} introduced the COmoving-LAgrangian enhanced particle mesh (COLA) scheme, where the large scale displacement is governed by the analytic calculation from second order Lagrangian theory (2LPT), and the particle mesh is used only to solve for the ``residual''  small scale displacement that affects the formation of halos. COLA scheme has gained a lot of attention recently \citep{cola13,scola15,lpicola2015,2016MNRAS.459.2327I,colahalo2015} because it adds a relatively small overhead to the vanilla particle mesh method, yet enforces the correct growth on large scales. As expected, this approach fails on small scales, where approximate tree solvers \citep{sunayama15} proposed to reduce the frequency of updating the small scale force produce better results (percentage level agreement at high $k$), although at a higher computational cost. An alternative approach to reduce the cost of small scale interactions is to neglect the Tree force calculation for particles that have formed halos \citep{khandai2009}.

The wall-clock time of a simulation is a somewhat overlooked issue (likely because the usual applications of approximate methods focus on a large number of independent mocks).  The wall-clock time is still of relevance from a practical point of view, especially so for certain applications. For example, for Markov Chains Monte Carlo or similar sampling algorithms, a shorter wall-clock time in the simulations allows more steps per chain which can be of significant importance. There are two possible ways to reduce the wall-clock time. The first is to redraw the compromise between amount of calculation and accuracy -- usually, less accurate results can be obtained by faster simulations. The second approach is to employ more computing resources. In the second approach the idea is that the implementation of the scheme can efficiently use a larger amount of computing resource -- the so called strong scaling performance. Strong-scaling will become even more relevant as the number of computing nodes of super-computing systems grows with time. Usually, the more complex is the code, the harder it is to enforce strong scaling, specially when moving to new platforms.

To address some of these issues, we implement a simple PM scheme into a new code we call FastPM, where the linear theory growth growth of displacement (Zel'dovich approximation or 1LPT, where nLPT stands for Lagrangian Perturbation Theory at order n) is enforced by choosing an appropriate set of kick and drift factors. The particle mesh solver in FastPM is written from scratch to ensure that it scales extremely well with a large number of computing nodes. We implement it both within our approach, and within the COLA implementation, so that we can compare their performances.

We will describe the FastPM code in Section 2 of this paper. We discuss the numerical schemes briefly in Section 2.1, then move on to discuss the 2-dimensional parallel decomposition in Section 2.2 and show the strong scaling of FastPM in Section 2.3. In Section 3, we explore the parameter space (number of time steps and force resolution) and investigate  favorable schemes for approximated halo formation.

\section{The FastPM code}

\subsection{Numerical Schemes}
The time integration in FastPM follows the Kick-Drift-Kick symplectic scheme described in \cite{quinn97}, but is modified to achieve the correct large scale growth. This is discussed further below. Here we discuss the transfer functions for force calculation in Fourier space. 

In a particle mesh solver, the gravitational force is calculated via Fourier transforms. First, the particles are painted to the density mesh, with a given a window function $W(r)$. We use a linear window of unity size \citep[cloud-in-cell]{Hockney:1988:CSU:62815}. We then apply a Fourier transform to obtain the over-density field $\delta_k$. A force transfer function relates $\delta_k$ to the force field in Fourier space.

There are various ways to write down the force transfer function $\nabla \nabla^{-2}$ in a discrete Fourier-space. The topic has been explored extensively by \cite{Hockney:1988:CSU:62815} in the context of high resolution PM and PPPM simulations. FastPM supports several combinations, which we describe below.
\begin{itemize}
\item Naive: naive Green's function kernels ($k^{-2}$) and differentiation kernels ($i \bf{k}$). This is used in COLA to solve for the residual motion with PM.
\begin{equation}
\mathbf{\nabla} \nabla^{-2} = i\mathbf{k}~
k^{-2}.
\end{equation}
\item H-E: Sharpening Naive scheme by de-convolving the CIC mass assignment window. This is an extremely popular set of choice because it is shown to minimize the small scale error in force calculation,
\begin{equation}
\mathbf{\nabla} \nabla^{-2} = i\mathbf{k}~
k^{-2} W^{-2}_\mathrm{CIC}(k),
\end{equation}
where $W_\mathrm{CIC} = \Pi_{d=x,y,z}\mathrm{sinc}^2{\frac{k_i L}{2N}}$.
\item Finite differentiation kernel (FastPM). The finite differentiation operator on a mesh is used, and no correction for mass assignment is applied. The 3-point second-order central differentiation operator in Fourier space is  \citep[also mentioned in][]{Hockney:1988:CSU:62815}:
\begin{equation}
\nabla^{-2} =  
\left(\sum_{d=x,y,z} 
\left(x_0 \omega_d \mathrm{sinc} \frac{\omega_d}{2}
\right)^2
\right)^{-1},
\end{equation}
where
$x_0 = L / N_g$ is the mesh size, and $\omega = k x_0$ is the circular frequency that goes from $(-\pi, \pi]$, and
\begin{equation}
\nabla = D_1(\omega) =  \frac{1}{6}
\left(
8 \sin \omega
- \sin 2\omega
\right).
\end{equation}
$D_1$ is derived in \cite{hamming89}. 
The $D_1$ filter is exactly the same as the 4-point discrete differentiation filter used in GADGET \citep{gadget2}, except we apply the function in Fourier space to simplify the implementation.

\item Gadget: naive Green's function kernels ($k^{-2}$), and with $D_1$, applying sharpening corrections for mass assignment. This is the kernel used in Gadget for the PM part of the force.
\begin{equation}
\mathbf{\nabla} \nabla^{-2} = i D_1(k) ~
k^{-2} W^{-2}_\mathrm{CIC}(k).
\end{equation}
\end{itemize}

In solvers (e.g. TreePM) where the short range force is calculated separately, a Gaussian function $W_g(k) = \exp [-(k r_\mathrm{s})^2]$ is often added to apodize the long range force and suppress grid anisotropy, $r_\mathrm{s}$ being the smoothing scale of long range force (typical slightly larger than the force mesh cell sizes) \citep{treepm2002,gadget2,habib13}. It is unclear this is necessary for pure particle mesh simulations where there is no short range force: some authors have completely neglected the smoothing filter, yet still obtained reasonable results \citep{cola13}; authors in other fields have suggested that a high order Gaussian smoothing performs well for certain problems \citep{2007JCoPh.226..379H}. We find that even a mild smoothing, with $r_s = L / N_m$ of the mesh resolution, significantly reduces the number of halos at the low mass end; we will discuss the effect due to choice of schemes in Appendix \ref{app:kernel}. For consistency in the main text paper we use the FastPM finite differentiation scheme without smoothing, which as shown in Appendix \ref{app:kernel} gives a sharper density field on larger scales, even though the kernels are less accurate on small scales in a traditional sense \citep{Hockney:1988:CSU:62815}.

FastPM accepts an arbitrary list of time steps. Shortcuts for several commonly used time stepping schemes (including linear and logarithmic) are provided:
\begin{itemize}
\item Time steps that are linear in scaling factor $a$. Linear stepping improves halo mass function, but requires assistance for an accurate linear scale growth factor if number of time steps is small \citep{cola13,scola15,lpicola2015,2016MNRAS.459.2327I}.
\item Time steps that are linear in $\log a$. Logarithmic stepping improves growth on large (linear) scales at the cost of underestimating small scales and the halo mass function \citep{qpm14}.
\item A hybrid scheme that controls the time step resolution in high redshift and low redshift independently:
\begin{equation}
(\delta a / a)^{-1} = \sqrt{(1/a_1)^2 + (a/a_2)^2)},
\end{equation}
where $a_1$ controls the early time stepping and $a_2$ controls the late time stepping. An example value of $a_1 = 0.05$ and $a_2 = 0.025$ gives about 90 steps from $a=0.025$ to $a=1.0$\citep{carlson09}.
\end{itemize}
In this paper, we will focus on linear $a$ stepping.

To compensate for the lack of short range resolution, the resolution of the force mesh can be boosted by a factor $B=N_m/N_g$. The size per side of the mesh ($N_m$) used for force calculation is $B$ times the number of particles per side ($N_g$). Most recent authors of PM/COLA advocated a mesh of $B=3$ in order to capture the non-linear formation of halos. \citep[][]{cola13,2016MNRAS.459.2327I}, while others advocated for $B=1$\citep[][]{lpicola2015}.  In FastPM $B$ can be provided as a function of time.  We will investigate the choice of $B$ in later in this paper.

\subsection{Memory Usage}
FastPM implements both COLA and vanilla PM. COLA requires storing the two 2LPT displacement fields $s_1$ and $s_2$, inducing a memory overhead relative to vanilla PM.
The state vector of a single particle in FastPM contains the position, velocity and acceleration, in addition, the initial position is encoded into an integer for uniquely identifying the particles. When COLA is employed, we also store the $s_1$ and $s_2$ terms. Currently, the position of particles $x$ is stored in double precision to reduce systematic evolution of round-off errors near the edge of the boxes, which can be concerning if stored in single precision. One can also store the relative displacement from the original position in a single precision, which can further reduce the memory consumption by a few percent.

The memory usage on the state vector is summarized in Table \ref{tab:statevector}. The memory usage per particle is 56 bytes for vanilla PM and 80 bytes for COLA. To account for load imbalance and particles near domain surfaces (ghosts), we always over-allocate the memory storage for particles by a factor of $A > 1$. Therefore, the total memory usage for storing the state vector is $M_1 = 56 A N_g^3$ for PM, and $M_1' = 80 A N_g^3$ for COLA. 

To avoid repeated conversion from particles to the mesh, FastPM creates two copies of the force mesh. The memory usage for the force mesh is $M_2 = 2 \times 4 (B N_g)^3 = 8 B^3 N_g^3$. Note that the buffer for domain decomposition and for creation of a snapshot is overlapped with the force mesh, and it thus does not incur further memory allocation.

In summary, the total memory usage of FastPM is 
\begin{equation}
M = M_1 + M_2 = \begin{cases}
(56 A + 8 B^3) N_g^3 & \text{PM}, \\
(80 A + 8 B^3) N_g^3 & \text{COLA}.
\end{cases}
\end{equation}

$A$ is typically bound by $1 < A < 2$. Therefore, the additional memory cost of COLA relative to vanilla PM to store $s_1$ and $s_2$ is: $37\% \sim 40\%$ for $B=1$, $20\% \sim 27\%$ for $B=2$, and $9\% \sim 15\%$ for $B=3$.


\begin{table}
\begin{tabular}{ccc}
Column & Data Type & Width (Bytes) \\
\hline
$x$ & double & 24 \\
$v$ & single  & 12 \\
$a$ & single  & 12 \\
$q$ & integer & 8 \\
$s_1$ & single & 12 \\
$s_2$ & single & 12 \\
\hline
Total PM & & $56$ \\
Total COLA & & $80$\\
\hline
\end{tabular}
\caption{Memory for the state vector, per particle.}
\label{tab:statevector}
\end{table}

\subsection{Domain Decomposition}
The domain decomposition in FastPM is 2-dimensional. Decomposing in 2 dimensions (resulting 1-dimension 'pencils' or 'stencils') is an effective way to deploy large Fourier transforms on a massively parallel scale \citep[see e.g.][]{pfft13,p3dfft}. Some gravity solvers implement a new 2-dimensional Fourier transform library \citep[e.g.][]{habib13}. We choose to reuse the publicly available implementation, {\small PFFT} by \cite{pfft13} for its minimal design. We note that {\small PFFT} was also used to improve the scaling of {\small P-GADGET} by \cite{feng15}.

FastPM decomposes the particles into the same 2-dimensional spatial domains of the real space Fourier transform mesh. In general, there are more mesh cells than number of particles (when $B > 1$), and it is therefore more efficient to create ghost particles on the boundary of a domain than creating ghost cells. We do not further divide the domain along the third dimension because that would induce further communication costs.

For large scale computations, a 2-dimensional decomposition has two advantages over 1-dimensional decomposition: 1) a smaller total surface area; and 2) a more balanced load. The surface area is directly proportional to the amount of communication for ghost particles. The surface area increases linearly with the number of processes ($O[P]$) for 1-dimensional decomposition; while for 2-dimensional decomposition, the scaling is close to $O[P^{1/2}]$. The unit of parallelism in 1-dimensional decomposition is a slab of $1\times N_g \times N_g$. Therefore, when the number of processes is greater than the number of slabs ($P > N_g$), the Fourier transform becomes extremely imbalanced. The unit of parallelism in 2-dimensional decomposition is a pencil of $1\times 1 \times N_g$, and the constraint is $P > N_g^2$. For a typical mesh size of 8,192 we used, the limit translates to $8,192^2=67,108,864$ processors for 2-d decomposition, a limit cannot be reached even with the next generation exa-scale facilities.

We point out that FastPM is not the first N-body solver implementing a 2-dimensional domain decomposing scheme. Previous implementations \citep[e.g.][]{habib13, feng15} mostly focused on weak-scaling of simulations to a large number of particles. The strong scaling of schemes that resolves galactic scale interaction (gravity, hydrodynamics, and feedback) typically suffer from the heavy imbalance as the average volume of a domain decreases, and requires over-decomposition of domains \citep{gadget2,menon14}. FastPM does not usually suffer from the over-decomposition since it does not implement any small scale interaction, as we will discuss in the next section.

\subsection{Strong Scaling of FastPM}
We run a series of tests to demonstrate the strong scaling performance of FastPM on the Cray XC-40 system Cori at National Energy Research Supercomputing Center (NERSC). The test simulation employs $1024^3$ (1 billion) particles in a box of $1024h^{-1}\unit{Mpc}$ per side, and a resolution of $N_s=40$/$B=3$. The cosmology used is compatible to the WMAP 9 year data.

The minimal number of computing nodes to run the simulation (due to memory constrains) is 4, which translates to $N_p = 128$ computing cores. We then scale up the computing scale all the way up to $N_p = 8,192$ (a factor of 64) computing cores. We measure the time spent in force calculation, domain decomposition and the generation of the 2LPT initial condition. Time spent in file operations (IO) is not shown, since it follows closely to the status of the file system instead of the scaling of the code. We note however, that the IO backend of FastPM (\textrm{bigfile}) is capable of achieving the peak performance of the file system in the BlueTides simulation \citep{feng15}.

\begin{figure*}
\includegraphics[width=0.9\columnwidth]{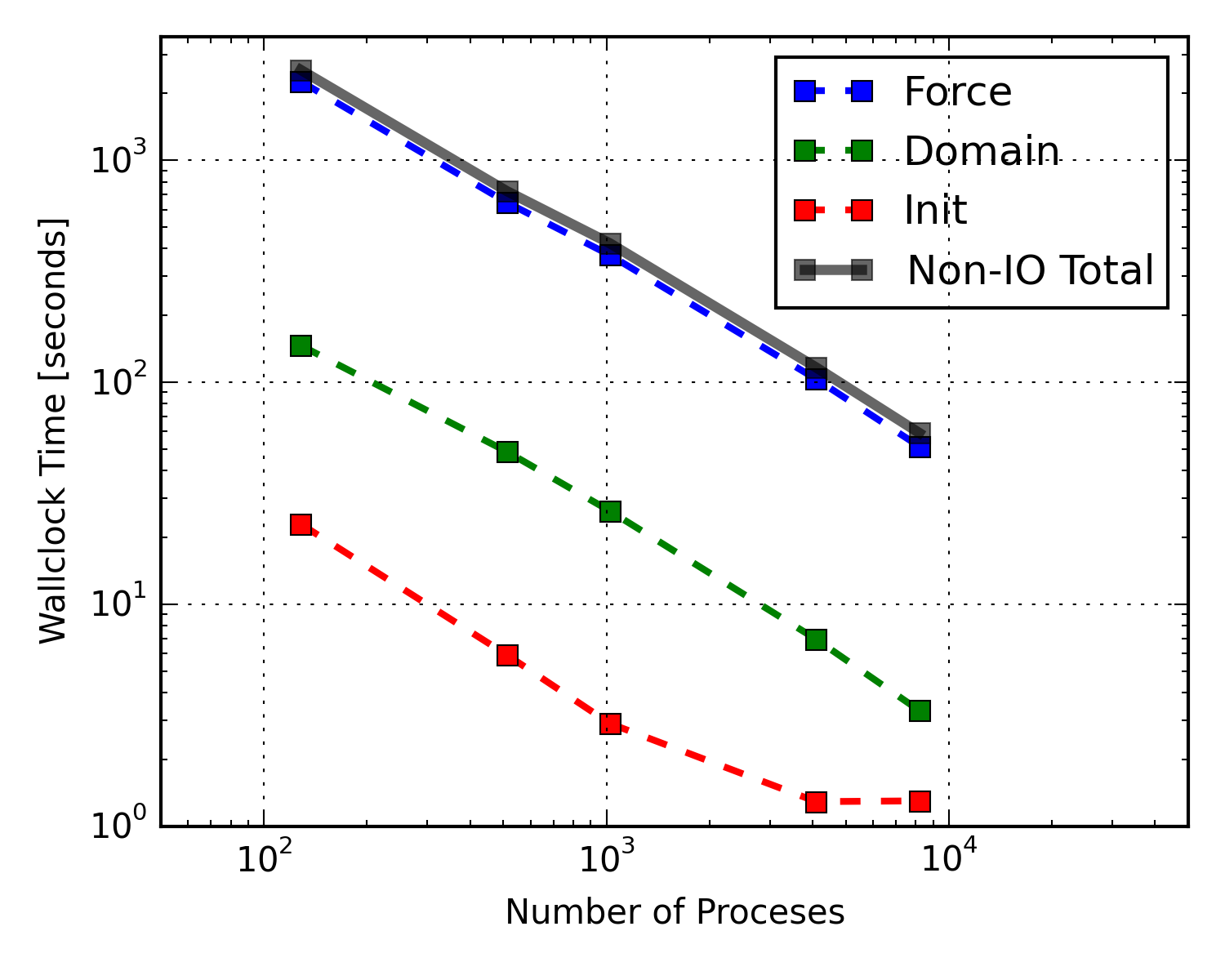}
\includegraphics[width=0.9\columnwidth]{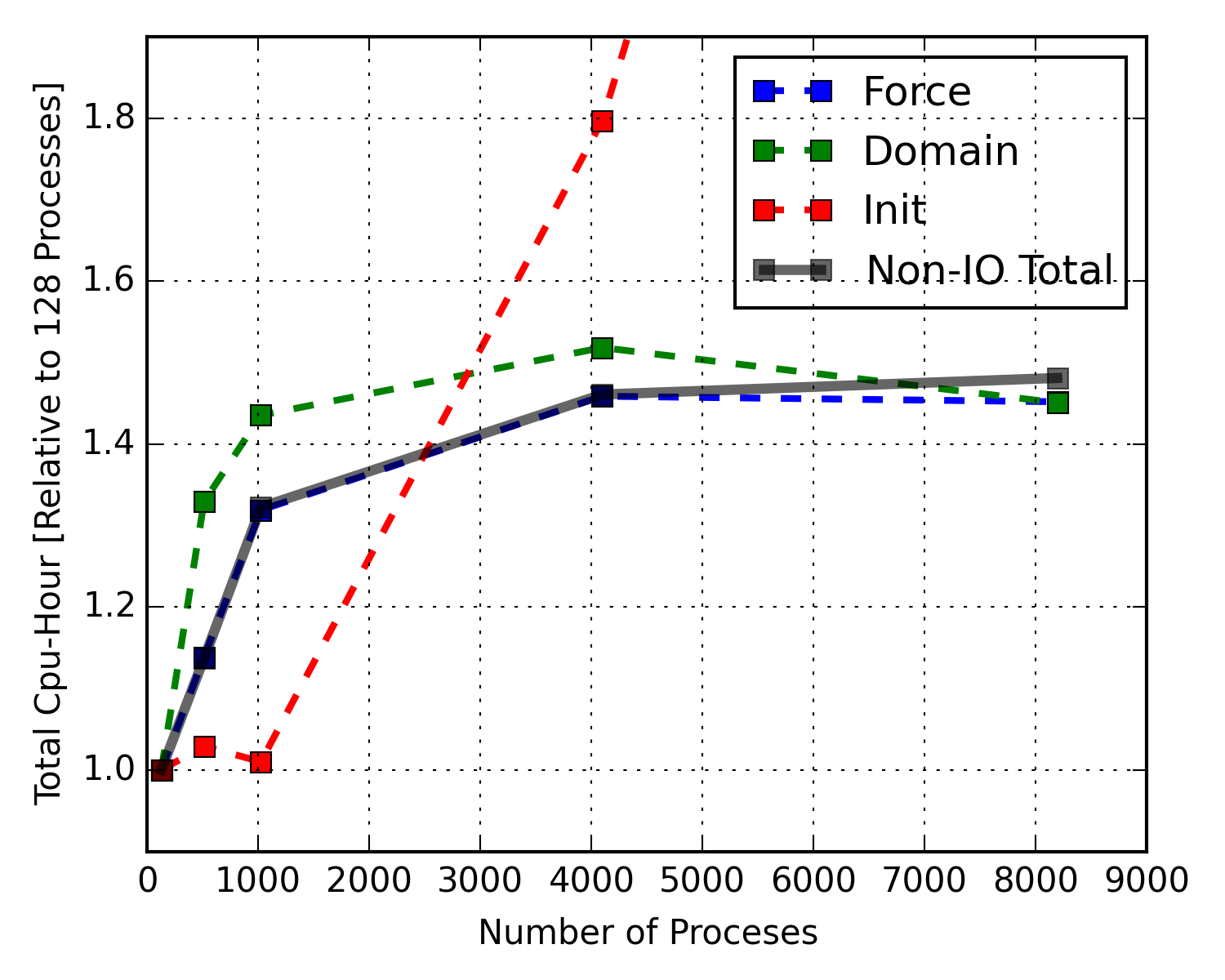}
\caption{Strong scaling of FastPM. We perform the test on a $1024h^{-1}\unit{Mpc}$ per side box on a $1024^3$ particle grid. We run a total of 87 time steps (blendspace time stepping configuration) with the $B=1, 2, 3$ force resolution scheme. We choose the variable force resolution scheme because in a traditional slab based particle mesh solver, the scaling would have stopped with 1024 cores for the $B=1$ mesh. The three components shown are 1) Force: the time spent in obtaining acceleration of particles; 2) Domain: the time spent in migrating particles between different processors; 3) Init; the time spent in generating the 2LPT initial condition.}
\label{fig:scaling}
\end{figure*}

We report the results of the scaling tests in Figure \ref{fig:scaling}. In the left panel, we see that the scaling of the total wall-clock time is close to the ideal $1 / N_p$ law. The scaling of 2LPT initial condition hits a plateau when more than 4096 computing cores are used, but this is not of particular concern since the fraction to the total only amounts 3\% percent even for the 8,192 core run. 

The deviation from the $1 / N_p$ law is shown in the right panel of Figure \ref{fig:scaling}. We show the evolution of the total CPU-hours (cost) as we increase the number of cores. For ideal $1 / N_p$ scaling, the line should be flat. We see that the total cost increases slowly with the number of cores, by a factor of less than 1.45 when the number of cores increased by a factor of 64. The increase of total cost is primarily due to the imbalance of particles that is associated with the decrease of volume per domain. For example, with the 8,192 core run, the most loaded domain handles 3 times of the average number of particles per domain. 
In a numerical scheme where small scale force is fully resolved (TreePM), the computing time increases very quickly with the growth of over-density, and this imbalance would have significantly increased the cost. However, in an approximated particle mesh solver (like FastPM), the dominating cost component is Fourier transform, the load of which is balanced relatively well thank to the 2-dimensional decomposition. The increase in synchronization time due to particle imbalance only slowly increase the total computing time. We point out that the relatively quick increase in cost at small number of cores ($N_p < 1,024$) is correlated with the crossing of communication boundaries of ``local''/ per-cabinet network on the Cray XC 40 system Cori where the test is performed.

OpenMP threading is usually invoked as a workaround for strong-scaling limitations \citep[e.g.][]{feng15}. For FastPM this particular context is no longer relevant. In fact, running with multiple threads is always slower than running with equal number of processes, due to the lack of thread parallelism in the transpose phase of the parallel Fourier transform\footnote{Refer to \url{https://github.com/mpip/pfft/issues/6} for some discussions of this issue.}. Therefore on current computer architectures, we do not recommend using more than 1 OpenMP thread in FastPM.

\subsection{Time Stepping}

\begin{figure}
\includegraphics[width=\columnwidth]{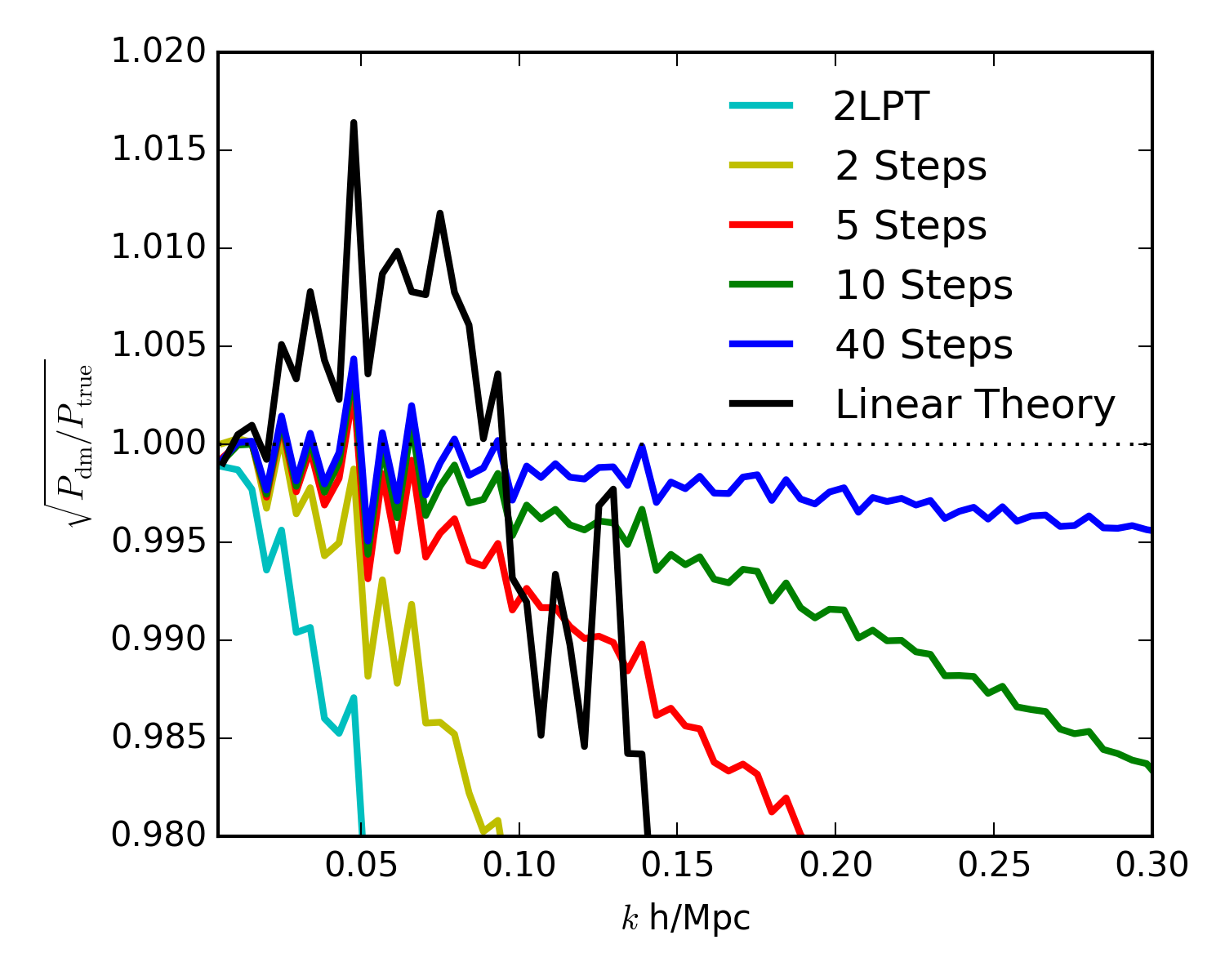}
\caption{Recovery of linear growth with FastPM. We show the recovery rate $\epsilon(k)$ at quasi-linear scales, divided by full N-body (RunPB). Cyan: 2LPT; Yellow 2 steps; Red: 5 steps; Green: 10 steps; Blue: 40 steps. The fluctuations are due to the sampling variance errors in recovering the initial condition (see text). We also show the ratio of linear modes to full N-body, which show that linear theory deviates from the nonlinear results over most of the range.}
\label{fig:bbc}
\end{figure}

When the number of time steps is limited, the Kick-Drift-Kick integration scheme fails to produce the correct growth on large scales \citep[e.g.][]{2016JCAP...04..047S,2016MNRAS.459.2327I}. 
We note that even very large scales are not fully linear. As seen in figure \ref{fig:bbc}, at $k_{th} \sim 0.03 h/\unit{Mpc}$, the linear theory model introduces an average 0.5\% systematic error in mode amplitude at $z=0$ due to nonlinear coupling of modes \citep[see also][]{foreman15,baldauf15,seljak1501,2008MNRAS.389.1675T}. 
The linear scale growth error is especially severe with linear $a$ time stepping, where the relative change in the growth factor can be large-- for example, if the first two time steps are at from $a=0.1$ ($z=9$) to $a=0.2$ ($z=4$).
Our numerical experiment shows that the error is already 1.5\% with a single step from $a=0.1$ to $a=0.2$.

One way of eliminating this error without substantially increase the number of time steps is to insist the large scale growth follows a model. For example, the COmoving-LAgrangian (COLA) particle mesh scheme calculates the large scale trajectory of particles with second order Lagrangian Perturbation theory \citep{cola13}, which yields an accurate growth of the large scale modes. The disadvantages are higher memory requirement and the need to split the force into 2LPT and PM minus 2LPT, which requires tuning that is somewhat dependent on the number of time steps. 

In a pure PM scheme the error is due to the incorrect assumption that force and velocity remains constant during the course of a time step. This assumption is violated even in the linear regime. It can be compensated with a set of modified discrete drift and kick factors, motivated by the Zel'dovich (ZA) equation of motion. We derive next these factors (denoted with subscript FASTPM).

To see this, we first write down the Zel'dovich equation of motion to first order,
\begin{eqnarray}
x_\mathrm{ZA} (a) &=& q + D(a) s_1 \\
p_\mathrm{ZA} (a) &=& a^2 \frac{dx_\mathrm{ZA}}{dt} = a^2 \frac{d D}{d a} \frac{da}{dt} s_1 \\
                  &=& \frac{dD}{d a} a^3 E(a) s_1 , \\
                  &=& a^3 E(a) g_p(a) s_1 \\
f_\mathrm{ZA} (a) &=& a \frac{dp_\mathrm{ZA}}{dt} = a^2 E(a) \frac{dp_\mathrm{ZA}}{da} \\
                  &=& a^2 E(a) \frac{d [a^3 E g_p(a)]}{da} s_1 \\
                  &=& a^2 E(a) g_f(a) s_1 .
\end{eqnarray}
We have followed the convention that $p = a^2 dx / dt$, $f = a dp / dt$, and $E(a) = H(a) / H(a=1)$ is the dimensionless Hubble parameter.
For simplicity we define the factors $g_p$ and $g_f$ as
\begin{eqnarray}
 g_p(a) &=& \frac{dD}{da} \\
 g_f(a) &=& \frac{d (a^3 E g_p)}{da} \\
        &=& \frac{d (a^3 E)}{da} \frac{d D}{d a} 
              + \frac{d^2 D}{da^2} a^3 E(a) 
             \\
        &=& 3 a^2 E \frac{d D}{d a} 
                     + a^3 \frac{d E}{d a} \frac{d D}{d a} 
                     + \frac{d^2 D}{da^2} a^3 E,
\end{eqnarray}
and their integrals as
\begin{eqnarray}
G_p(a) &=& D(a) \\
G_f(a) &=& a^3 E(a) g_p(a)
\end{eqnarray}
Next, we rearrange the ZA equation of motion as drift/kick operators by integrating ZA over the time step and eliminating the ZA displacement $s_1$ from the equation of motion,
\begin{eqnarray}
\Delta[x_\mathrm{ZA}]_{a_0}^{a_1} &=& \Delta[D(a)]_{a_0}^{a_1} s_1 \\
     &=& 
     \frac{p_\mathrm{ZA}(a_r)}{a_r^3 E(a_r)} 
      \frac{\Delta[G_p]_{a_0}^{a_1}}{g_p(a_r)} 
      \\
\Delta[p_\mathrm{ZA}]_{a_0}^{a_1} &=& 
    \frac{f_\mathrm{ZA}(a_r)}{a_r^2 E(a_r)} 
    \frac{\Delta[G_f]_{a_0}^{a_1}}{g_f(a_r)}  
\end{eqnarray}
We can now define the FastPM modified drift and kick factor
\begin{eqnarray}
\mathcal{D}_\mathrm{FASTPM} &= & 
\frac{\Delta[x_\mathrm{ZA}]_{a_0}^{a_1}}{p_\mathrm{ZA}} =
\frac{1}{a_r^3 E(a_r)} 
     \left(
      \frac{\Delta[G_p]_{a_0}^{a_1}}{g_p(a_r)} 
      \right) \\
\mathcal{K}_\mathrm{FASTPM} &= &
\frac{\Delta[p_\mathrm{ZA}]_{a_0}^{a_1}}{f_\mathrm{ZA}} = 
\frac{1}{a_r^2 E(a_r)} 
    \left(
        \frac{\Delta [G_f]_{a_0}^{a_1}}{g_f(a_r)} 
    \right).
\end{eqnarray}
These operators can be used to construct any finite integration steps, which integrate the equation of motion of ZA solution exactly. Note that we keep $a_r$ as the reference time for the step. In a standard Kick-Drift-Kick scheme, $a_r$ is $a_0$ at the first step, but otherwise $a_r$ is between $a_0$ and $a_1$.

It is trivial to show these factors converges to the usual drift and kick operators when the time steps are small. The usual drift and kick operators are \citep{quinn97}
\begin{eqnarray}
x(a_1) &=& x(a_0) + p(a_r) \int_{a_0}^{a_1} \frac{1}{a^3 E} da \\
p(a_1) &=& p(a_0) + f(a_r) \int_{a_0}^{a_1} \frac{1}{a^2 E} da,
\end{eqnarray}
of which, the Drift and Kick factors are
\begin{eqnarray}
\mathcal{D}_\mathrm{PM} &=& \int_{a_0}^{a_1} \frac{1}{a^3 E} da \\
\mathcal{K}_\mathrm{PM} &=& \int_{a_0}^{a_1} \frac{1}{a^2 E} da.
\end{eqnarray}
When $a_1 \to a_0$, by definition we have
\begin{equation}
\frac{\Delta G_p}{\Delta a} \to g_p, \frac{\Delta G_f}{\Delta a} \to g_f,
\end{equation}
and
\begin{equation}
\mathcal{D}_\mathrm{FASTPM}  \to  \mathcal{D}_\mathrm{PM}, 
\mathcal{K}_\mathrm{FASTPM}  \to  \mathcal{K}_\mathrm{PM} .
\end{equation}
Therefore, on infinitesimal time steps, the two sets of operators are identical.

In Figure \ref{fig:bbc}, we show the large scale power spectrum of FastPM divided by the full N-body simulation. We also show the comparison to the linear theory prediction, which deviates from nonlinear already at very low $k$. Even with 2 time steps, i.e. a single full time step beyond 2LPT, the modified scheme is able to match the nonlinear growth of a full N-body simulation at $k \sim 0.02h/Mpc$, a significant improvement over 2LPT or linear theory. In contrast, full N-body codes often do not match each other at very low $k$ \citep{2008CS&D....1a5003H,2016JCAP...04..047S}, with constant offsets between them, likely due to inaccuracies in the linear growth. We thus believe our modified Kick-Drift scheme could be useful even for standard high resolution simulations. 

We point out that it is possible (as we have done in the first version of this paper and FastPM) to calibrate the Kick or Drift factor against linear theory, or non-linear power spectrum measured from a low resolution PM simulation in the limit of many steps, which gives nearly identical results. Both time stepping schemes are also implemented in FastPM.

\section{Dark matter and halo benchmarks with FastPM}

In this section, we investigate the accuracy of FastPM against computational cost, varying several of its parameters. We focus particularly on halo formation with FastPM, but we also compare it against the dark matter statistics. The particular application we have in mind is to find an approximate halo formation scheme that can be used for a fast extraction of galaxy statistics. To be more specific, we would like to find an approximation scheme that reduces the cost (CPU time), while also reducing  the systematic error of the halo statistics to an acceptable level (defined more precisely below). 

The parameters we investigate are number of time steps $N_s$ and the force resolution $B=N_m / N_g$, the ratio between force mesh and number of particles per side. We also perform a comparison between (our version of) COLA and FastPM. The FastPM simulations used in this work are listed in Table \ref{tab:list-of-sims}.  

The simulations are compared against a TreePM simulation on $2048^3$ particles in a $1380 h^{-1} \unit{Mpc}$ per side box (RunPB). For the FastPM simulations, we reconstruct the $s_1$ and $s_2$ 2LPT terms from a single precision $z=75$ initial condition of RunPB, and extrapolate the initial displacement to $z=9.0$. This ensures that the FastPM simulations are using the same initial modes as RunPB, up to the numerical errors. In all simulations, the Friend-of-Friend finder uses a linking length of 0.2 times the mean separation of particles \citep{davis85}. Our results are robust against linking length, as the main purpose of linking length is to identify over density features and apply abundance matching. In Appendix \ref{app:linking}, we also compare the results of $N_s=10$/$B=2$ simulations against those from a shorter linking length of 0.168, which reduces the halo bridging effect of FoF.

\begin{figure}
\includegraphics[width=\columnwidth]{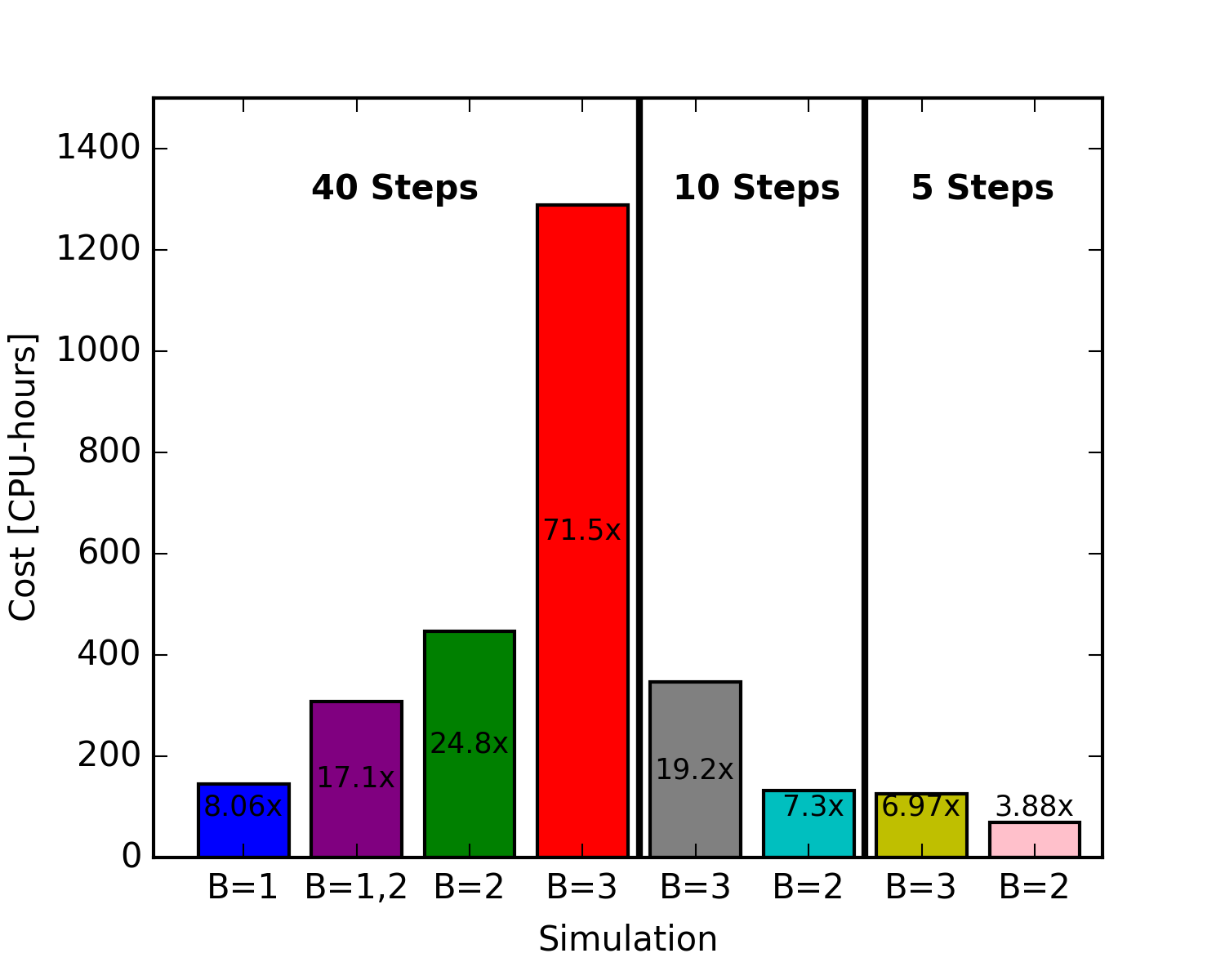}
\caption{The integrated CPU-hours. The text in each vertical bar shows the ratio to generating a 2LPT initial condition. COLA and PM use almost same amount of CPU-hours, thus are not independently shown.}
\label{fig:runtime}
\end{figure}

We list the total CPU time used in each run in Figure \ref{fig:runtime}. The total CPU time increases with $B$ and $N_s$. The most expensive scheme we considered is the $N_s=40$/$B=3$ scheme suggested by \cite{2016MNRAS.459.2327I}, using 1300 CPU hours each on the reference system (70 times of 2LPT). Reducing the number of time steps and the force resolution can drastically reduce the computing time. For example, while $N_s=40/B=3$ costs 72 times 2LPT, our favorite scheme $N_s=10$/$B=2$ uses only 7 times of cost of 2LPT, while $N_s=5$/$B=2$ reduces this to 4. 

\begin{table}
\centering\begin{tabular}{ccc}
\hline
\hline
Model & Number of Steps & Force Resolution \\
\hline
COLA & 5     & 3 \\
PM   & 5     & 3 \\
COLA & 10     & 3 \\
PM   & 10    & 3 \\
COLA & 10     & 2 \\
PM   & 10    & 2 \\
COLA & 20     & 3 \\
PM   & 20    & 3 \\
COLA & 40     & 3 \\
PM   & 40    & 3 \\
\hline
\hline
\end{tabular}
\caption{List of Simulations.}
\label{tab:list-of-sims}
\end{table}

\subsection{Definitions of benchmarks}

We use the ratio of mass function  $\phi_1(M) / \phi_{2}(M)$ to illustrate the difference in mass function. \footnote{1 stands for the approximated model and 2 stands for the accurate model.} We use abundance matching to correct for the difference in mass function, but note that more complicated alternatives have been used by other authors as well \citep[e.g.][]{sunayama15,2016MNRAS.459.2327I}.

We define the transfer function $T$ as the square root of the ratio of the power spectra
\begin{equation}
T(k, \mu) = \sqrt{P_1(k, \mu) / P_{2}(k, \mu)} .
\end{equation} 
The agreement is defined as good when $T$ is close to 1. The transfer function measures the relative bias between two fields. 

Note that transfer function is often defined as the ratio of the 
cross-power spectrum to auto-power spectrum: the two definitions are the same if 
the cross correlation coefficient is unity. 
We define the cross correlation coefficient $r$ as
\begin{equation}
r(k, \mu) = P_{1,2}(k, \mu) / \sqrt{P_1(k, \mu) P_{2}(k, \mu)},
\end{equation}
where $P_{1, 2}$ is the cross power between the approximated and the accurate model. For halos, we always use the catalog after abundance matching them. We do this by rank ordering them by assigned halo mass, and then selecting the same number of the most massive halos for the two catalogs. 

\begin{figure}
\includegraphics[width=\columnwidth]{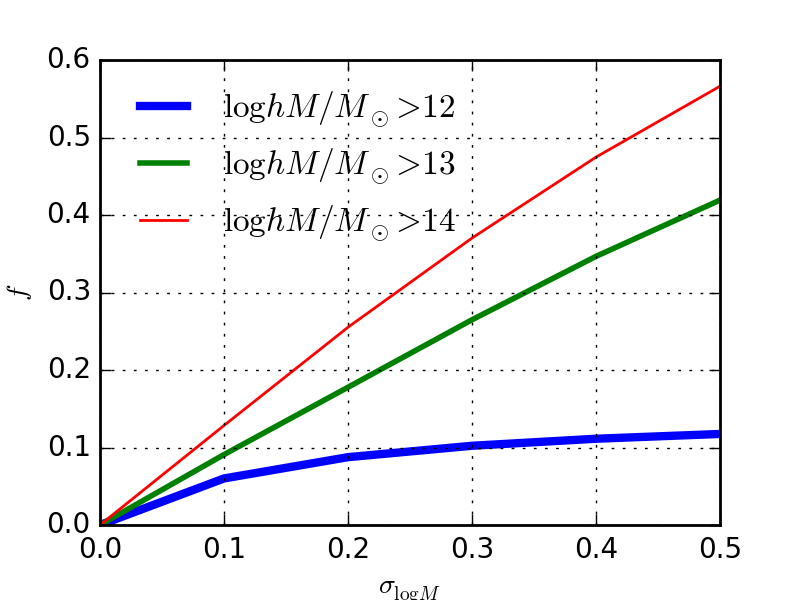}
\caption{Stochasticity induced by mass scattering. We measure the stochasticity (large scale) due to adding mass scattering (given by $\sigma$) to the halo catalog. Three different abundance cut, corresponds to $M=10^{12, 13, 14} h^{-1} M_\odot$ are shown in colors. The stochasticity increases with the mass scattering, which we identify as the main source of scale independent stochasticity. (See text)}
\label{fig:scatter}
\end{figure}
From this we can define another related quantity, the dimensionless stochasticity, 
\begin{multline}
   f(k, \mu) = 
   \sqrt{(1 - n P_1(k, \mu))(1 - n P_2(k, \mu))} \\
   + (1 - n P_{1,2}(k, \mu)),
\end{multline}
where $n$ is the number density of halos, which is identical in two simulations due to the abundance matching. The stochasticity is 1 when two catalogs contains completely different halos, and 0 when two catalogs are identical. So stochasticity $f$ expresses the fraction of misidentified halos in the approximate simulation. Typically 
this happens because the code has assigned an incorrect mass to the halo, so that the halo does not enter the abundance matched catalog. We could have also defined stochasticity using a mass error instead, but we do not pursue this here.
Our definition of stochasticity is $k$ dependent: if the halo positions are incorrect then we expect stochasticity $f$ to increase with $k$, before decreasing again to converge to the shot noise limit. 

While the numerical inaccuracies are one source of stochasticity, in practice we also have observational limitations. We typically observe galaxy luminosity, and select all the galaxies above a certain luminosity threshold (which can change with redshift). But there is a scatter between luminosity and halo mass and as a consequence a galaxy catalog does not correspond to the most massive halos at the same abundance (we are ignoring satellite galaxies in this discussion). To investigate this we take a given halo mass catalog, add scatter to it, and rerank the halos. We then determine stochasticity, i.e. the fraction of halos that are the same between this catalog and the original one, at the same abundance.

In Figure \ref{fig:scatter} we show the increase of stochasticity induced by increasing the mass scatter (in $\log M$) in the RunPB halo catalog. 
For example, the scatter of 0.23, as expected in BOSS CMASS catalog, corresponds to a stochasticity of 20\% for typical CMASS halo mass \citep{2015arXiv150906404R}. So as long as FastPM stochasticity is significantly smaller than this there is no need to make it exactly zero. 

It is worth pointing out that stochasticity or cross-correlation coefficient is in 
some sense the more important benchmark than the transfer function. This is because if the correlation coefficient is unity (or stochasticity zero) 
one can still recover the true simulation by 
multiplying the modes of the approximate simulation by the transfer function. 
Of course this requires the transfer function to be known, but it is possible that 
it is a simple function of paramaters, such that one can 
extract it from a small set of simulations 
without a major computational cost. In the following we will however explore 
all of the benchmarks defined here. These are summarized in 
table \ref{tab:benchmarks} for each simulation.
\begin{table*}
\begin{tabular}{ccccc}
\hline
\hline
Samples     & Transfer Function$^{\dagger}$ & Cross Correlation\newline Coefficient$^{\dagger}$ &Stochasticity$^{\dagger}$ & Mass Function \\
\hline
Matter     & Y & Y & N & N/A \\
$\log h^{-1} M / M_\odot \ge 12$  & Y & Y & Y & Y \\
$\log h^{-1} M / M_\odot \ge 13$ & Y & Y & Y & Y \\
$\log h^{-1} M / M_\odot \ge 14$ & Y & Y & Y & Y \\
\hline
\hline
\end{tabular}
\\
$^{\dagger}$ For these benchmarks, The halo mass in FastPM simulations are reassigned by abundance matching against halos in RunPB.
\caption{List of Benchmarks.}
\label{tab:benchmarks}

\end{table*}
\subsection{Varying Number of Time Steps}

In this section, we discuss the effects due to varying the number of time steps employed in the simulation $N_s$, while fixing the force resolution at $B = 3$.

\begin{figure}
\includegraphics[width=\columnwidth]{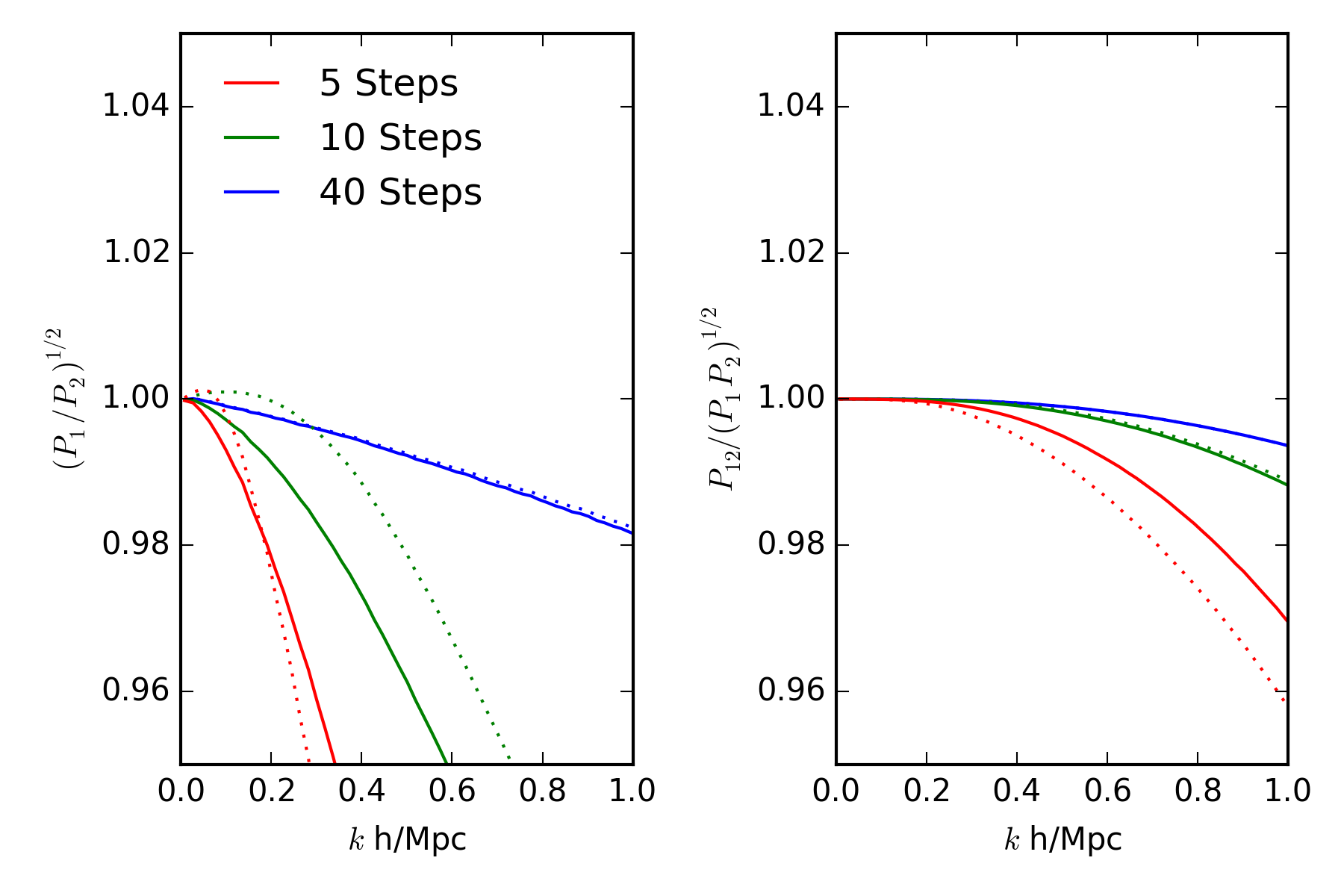}
\caption{Benchmarks on matter density, varying number of time steps. Left: transfer function. Right: cross correlation coefficient.
Colors:$N_s$ = 5 (red), 10(green), and 40(blue) and 200(cyan). Solid : PM. Dots : COLA. 200 Step run is with a lower resolution ($B=2$). The 200 step (cyan) and 40 step (blue) line has overlapped in the right panel.}
\label{fig:matter}
\end{figure}

\begin{figure}
\includegraphics[width=\columnwidth]{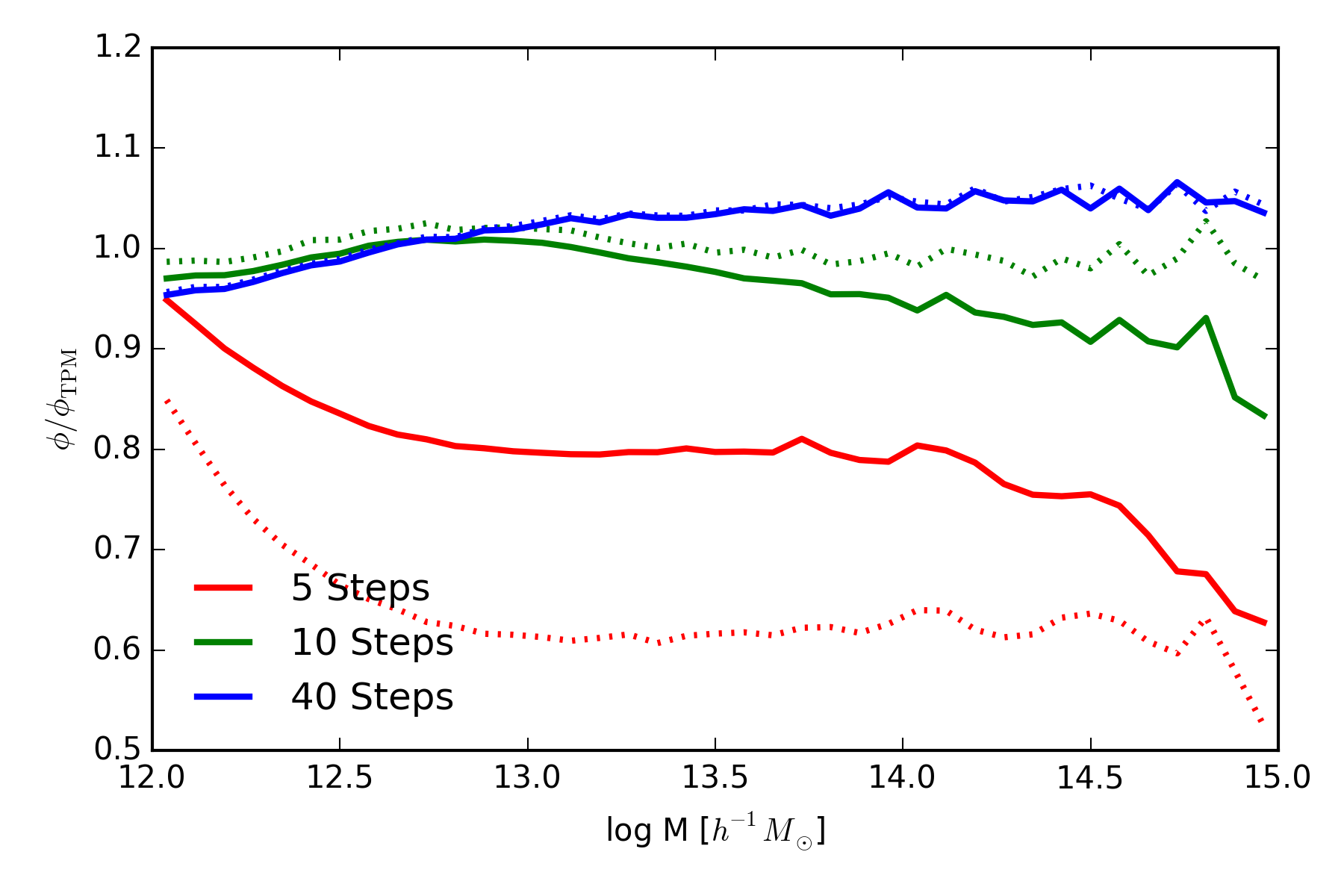}
\caption{Benchmark of halo mass function, varying number of time steps. We show the number of times steps (color: red, green, blue) from $N_s$ = 5, 10 and 40. Solid: PM. Dots: COLA.}
\label{fig:massfunction}
\end{figure}

The transfer function and cross correlation coefficient of matter density relative to the RunPB simulation is shown Figure \ref{fig:matter}. With 5 steps, the cross correlation coefficient is 93\% at $k = 1 h/\unit{Mpc}$. Increasing the number of time steps does improve transfer functions and cross correlation coefficients significantly. With 40 steps, the transfer function is close to 98\% at $k = 1 h/\unit{Mpc}$, and the cross correlation coefficient is close to 99\% at $k = 1 h/\unit{Mpc}$. These metrics indicate that if a high accuracy matter density field is of interest a 40 step simulation is preferred. It is worth pointing 
out that by extracting the transfer function, and then multiplying the modes with it,
one obtains nearly perfect results up to $k = 1 h/\unit{Mpc}$ even with 10 steps, given that the cross correlation coefficient is 99\% or larger over this range. 

We also observe that with the $N_s=10$ runs COLA gives a larger transfer function than FastPM at all scales, while $N_s=5$ COLA gives a smaller transfer function and a significantly worse cross correlation coefficients at all scales. It is worth noting that COLA has a free parameter $n_\mathrm{LPT}$ which needs to be tweaked for number of steps and cosmology parameters \citep{cola13}. In our tests we find that COLA is very sensitive to the choice of this parameter. In contrast, the kick and drift scheme in FastPM does not require tweaking of any parameters.

We next look at the halos. Before applying abundance matching, we first show the mass function relative to RunPB is shown in Figure \ref{fig:massfunction}. For runs with more than 10 steps, the mass function has converged to 10\% agreement with RunPB regardless of whether COLA or PM is used. However, if mass accuracy is required then 
5 steps appears to be insufficient, as the 5 step PM run recovers only 80\% of the mass function (60\% for COLA). This is not necessarily a problem, since exact mass assignment is not required in a galaxy survey with a given abundance, as long as the rank ordering is preserved. However, in practice lower number of steps also introduces errors in the mass assignment, which increase the stochasticity, as we show below. 

\begin{figure*}
\includegraphics[width=\textwidth]{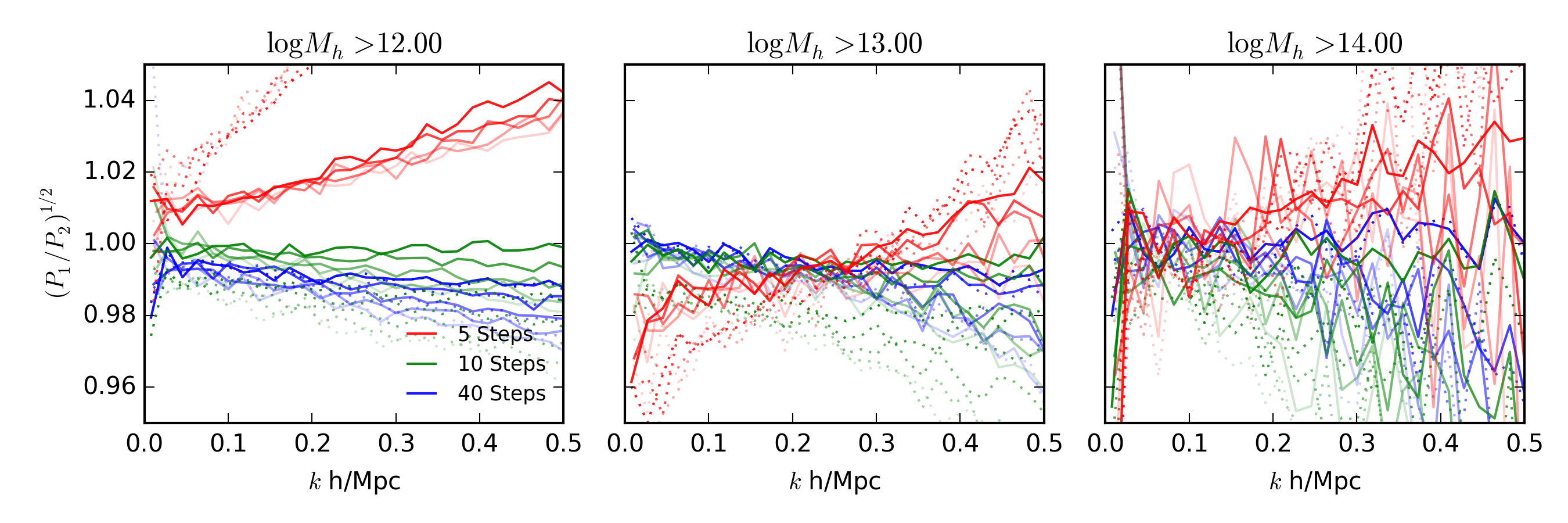}
\includegraphics[width=\textwidth]{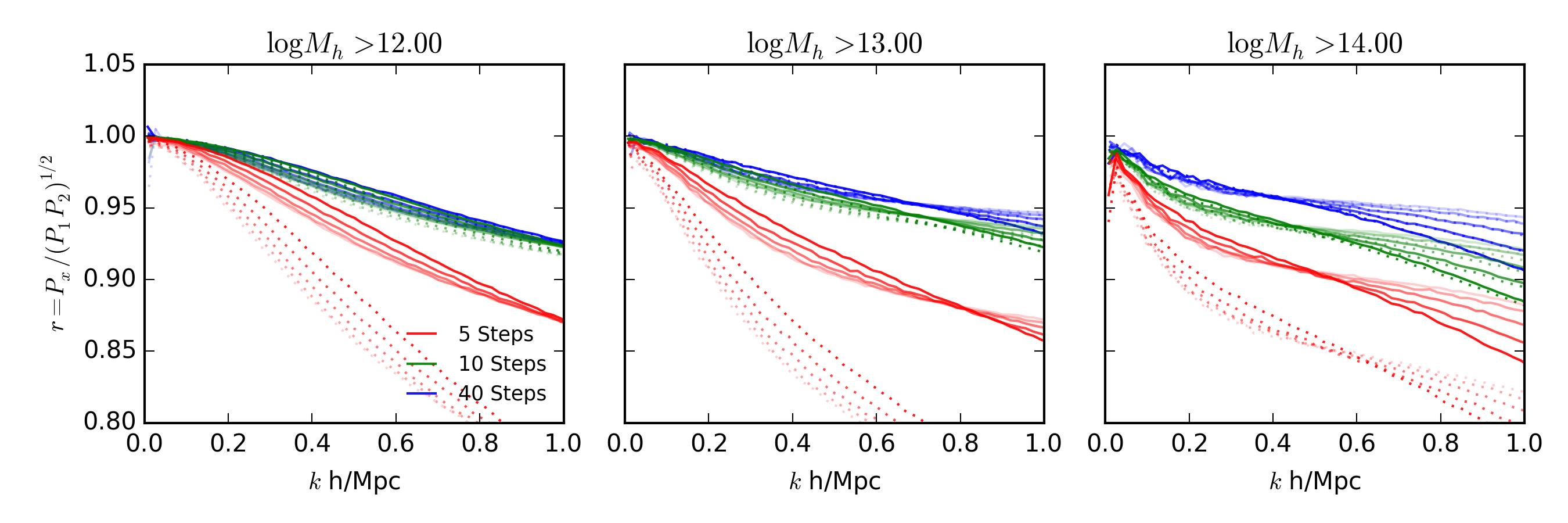}
\includegraphics[width=\textwidth]{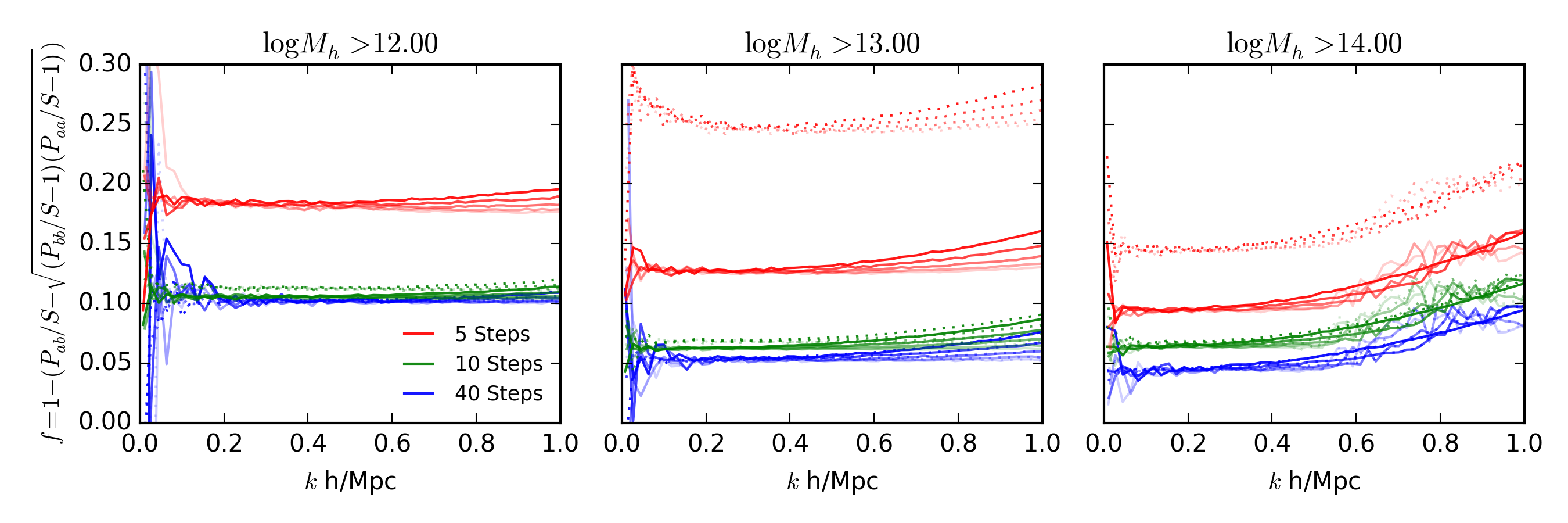}
\caption{Benchmarks on halos, varying number of time steps. Top panel: transfer function. Center panel: cross correlation coefficient. Bottom panel: stochasticity. 
Colors:$N_s$ = 5 (red), 10(green), and 40(blue). Opacity: Line of sight angle, $\mu$ = 0.1 (transparent), 0.3, 0.5, 0.7, 0.9 (opaque). Solid : PM. Dots : COLA. $S=1/n$ is the shot-noise level.}
\label{fig:benchmarks}
\end{figure*}

The rest of the benchmarks are calculated after applying abundance matching to reassign halo masses in the FastPM simulations. In Figure \ref{fig:benchmarks}, we show the benchmarks on the transfer function, cross correlation coefficient and stochasticity of halos as the total number of time steps $N_s$ is varied. Three mass threshold, $M = 10^{(12, 13, 14)} h^{-1} M_\odot$ are used. All the results are 
for $z=0$. 
We find the following results:

1) Beyond 10 steps the improvement is very limited. This is very different from the matter density field, where one gains major advantage using 40 steps. The additional steps therefore mostly improves the profile and velocity dispersion of halos.

2) For abundance matched halos, PM out-performs COLA at low number of time steps. This can be most clearly seen in the 5 step runs, where PM is nearly a factor of 2 closer to exact solution at all $k, \mu$ values. The PM advantage over COLA decreases as the number of time steps increases. At 40 steps, PM and COLA converges to the same result. COLA splits the displacement into a residual field and a large scale 2LPT component. It is worth noting that even though the 10 step COLA simulations give better matter transfer function than PM (as seen in Figure \ref{fig:matter}), the advantage in matter density seem to be consistently hurting the performance in halos. This could be because COLA has free parameters whose performance may have been optimized for the dark matter benchmarks. In contrast, there are no free parameters in FastPM. 

3) FastPM matches more massive halos better than less massive halos. For example, for the 10 step runs, the stochasticity reduces from 10\% at $M > 10^{12} h^{-1}M_\odot$ to 7\% at $M > 10^{14} h^{-1}M_\odot$. However, the overall stochasticity is given by $f/n$, and since $n$ is a lot higher for the low mass halos 
the absolute stochasticity is a lot lower for low mass halos, despite the larger value 
of $f$.
A 10\% stochasticity is likely more accurate than our current level understanding of the halo mass - galaxy luminosity relation \citep{behroozi2010,more2009,yang2009}.  For example, the scatter in the halo mass at a fixed luminosity is 0.4 dex for $10^{12} h^{-1} M_\odot$ halos \citep{behroozi2010}, if all uncertainties are considered, and even larger for the larger halo masses. As a comparison, for a scatter of 0.18 dex in halo mass, we introduce a stochasticity of $f \sim 0.10, 0.18, 0.22$ for halos of mass $10^{12,13,14} h^{-1} M_\odot$. Hence we believe that the 
stochasticity levels generated by $N_s=10$, or even $N_s=5$, suffice given the current observational uncertainties in halo mass determination. 

4) Redshift space statistics ($\mu>0$, where $\mu=k_{||}/k$ and $k_{||}$ is the component of the wavevector $k$ along the line of sight) are not very different from the real 
space statistics ($\mu=0$). Hence redshift space distortions do not significantly 
affect the conclusions above. 

5) Scale dependence of $f$ is weak for low mass halos, gradually increasing to higher masses. For the highest mass bin we observe it to increase by 0.05 to $k=1{\rm h/Mpc}$. We expect that a low resolution PM gets the halo centers wrong by a fraction of their virial radius: hence, for lower mass halos the absolute error is smaller.  

Based on these observations one does not need more than 10 steps if halos at $z=0$ are of interest, and indeed for many applications even a 5 step FastPM, possibly corrected with the transfer function, suffices. In addition, we comment that the requirement on the number of steps are similar at higher redshifts (we tested up to $z=1$). Since our steps are uniform in expansion factor $a$, a 10 time step FastPM simulation that runs to $z=0$ would naively have the effective performance of 5 steps at $z=1$, but we observe that actual performance is more in between 5 and 10 steps. 

\subsection{Varying Force Resolution $B$}

In this section, we discuss the effects due to varying the resolution of the force mesh employed in the simulation $B$. As we have shown $N_s=10$ is of sufficient accuracy for halos, so we will fix the number of time steps at $N_s = 10 $ in this section.

\begin{figure}
\centering
\includegraphics[width=\columnwidth]{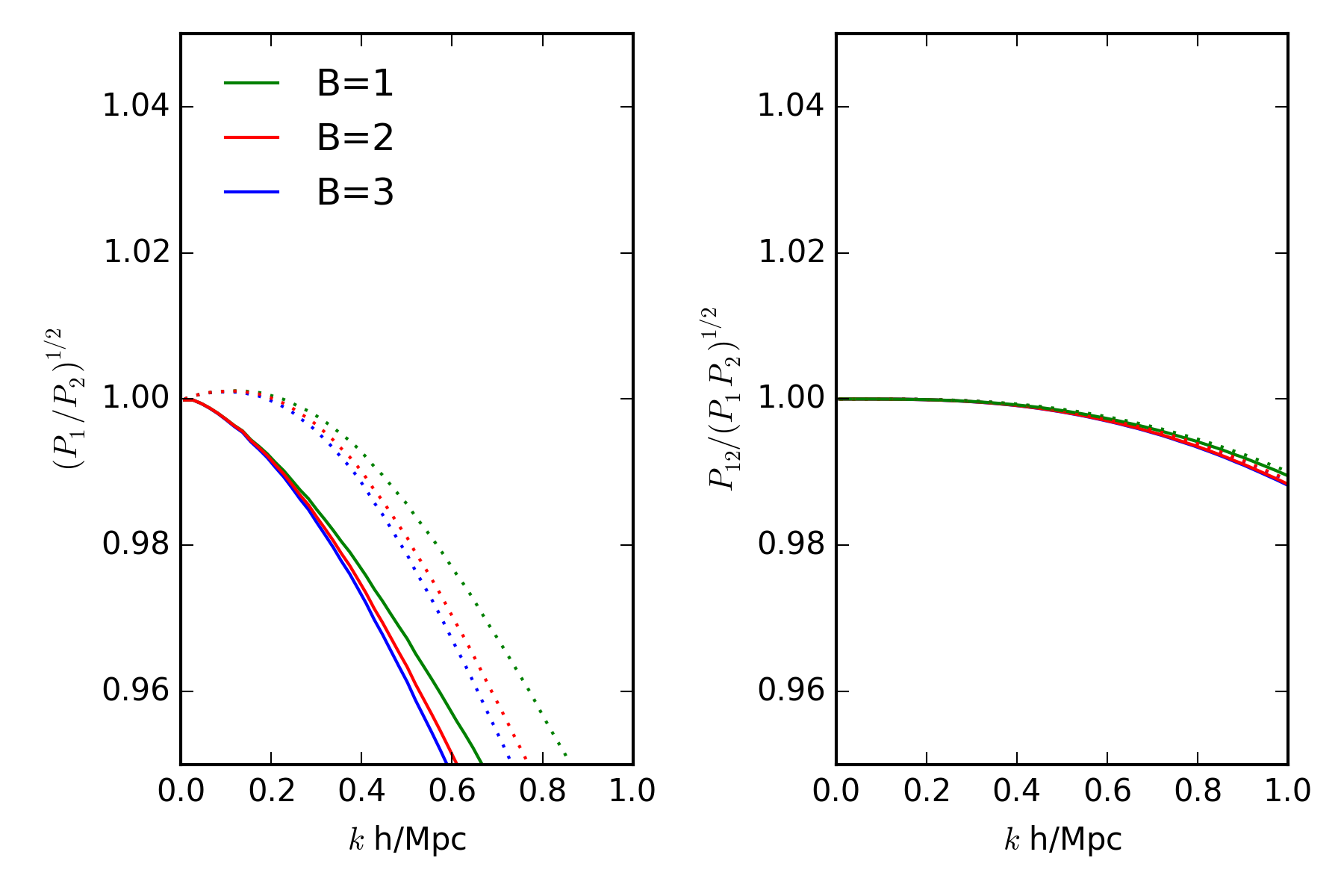}
\caption{Benchmarks on matter density, varying force resolution.  Red: $B=2$ ; Blue: $B=3$; Gray: $B=1$. Solid: PM; Dots: COLA. The number of steps is fixed to 10.}
\label{fig:matter-10}
\end{figure}

The transfer function and cross correlation coefficient of dark matter density relative to the RunPB simulation is shown Figure \ref{fig:matter-10}. We see that going from $B=3$ to $B=1$ the transfer function at $k < 1 {\rm h/Mpc}$ barely changes, and the cross correlation coefficient changes even less.

\begin{figure}
\includegraphics[width=\columnwidth]{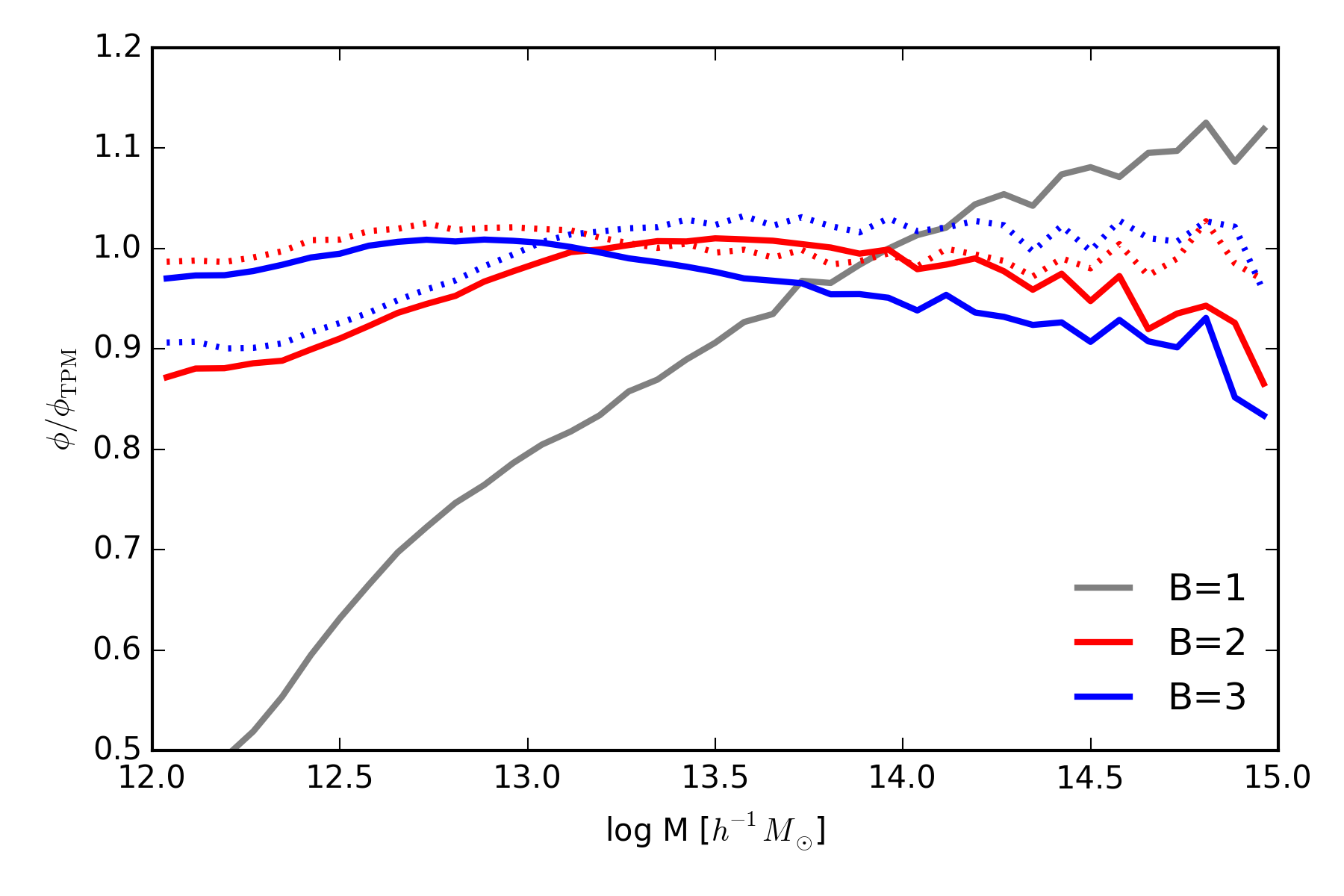}
\caption{Benchmark on halo mass function, varying force resolution. Green: $B=1$ ; Red: $B=2$ ; Blue: $B=3$. Solid: PM; Dots: COLA. The number of steps is fixed to 10.}
\label{fig:massfunction-10}
\end{figure}

As in the previous section, before applying abundance matching we investigate the mass function benchmark in Figure \ref{fig:massfunction-10}. We see that regardless of whether PM or COLA is employed, as long as $B \ge 2$, the mass function is recovered at 90\% level for $M > 10^{12} h^{-1} M_\odot$. However, with a lower resolution, $B = 1$, only 80\% of the halo mass function is recovered at $M = 10^{13} h^{-1} M_\odot$, and even fewer halos are found at lower masses. This indicates that due to the low resolution, the density contrast in $B=1$ is insufficient for detecting halos of $M \le 10^{13} h^{-1} M_\odot$, as also seen in \cite{2007ApJ...671.1160L}. However, one may still be able to salvage information about halos in the regime where Friend-of-Friend finder fails. For example, it is possible to combine a stochastic sampling method \citep[e.g. QPM or PATCHY][]{qpm14,patchy2014} for less massive halos, but 
this may increase stochasticity and move $f$ closer to unity. 

\begin{figure*}
\includegraphics[width=\textwidth]{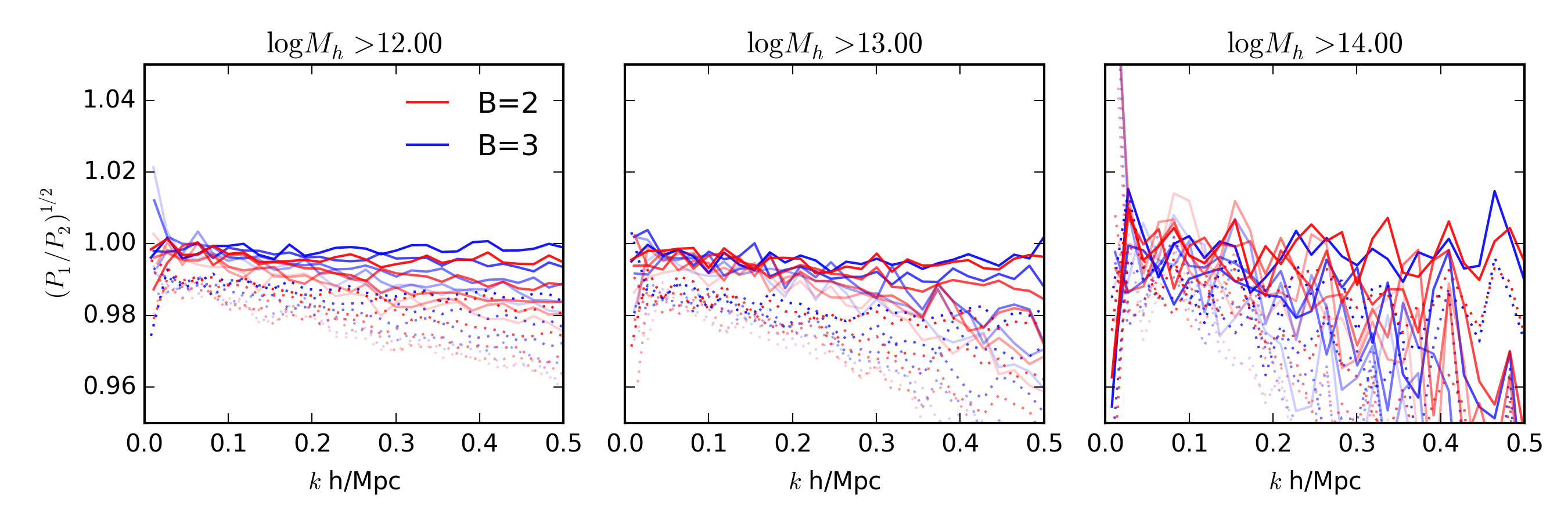}
\includegraphics[width=\textwidth]{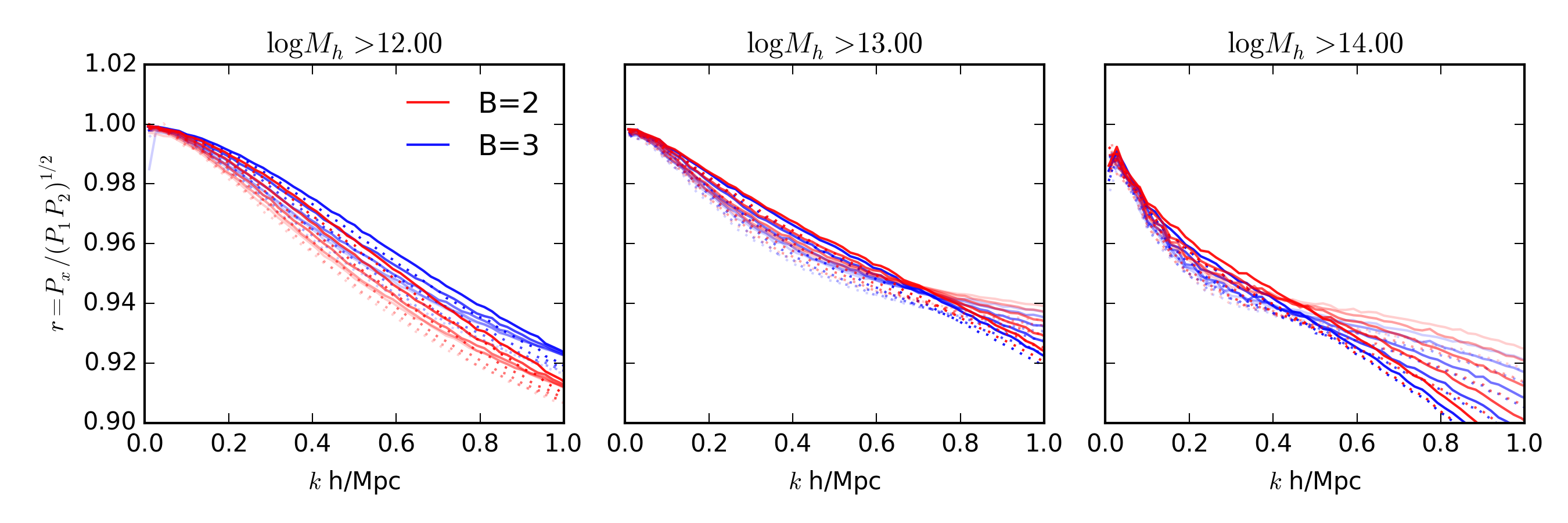}
\includegraphics[width=\textwidth]{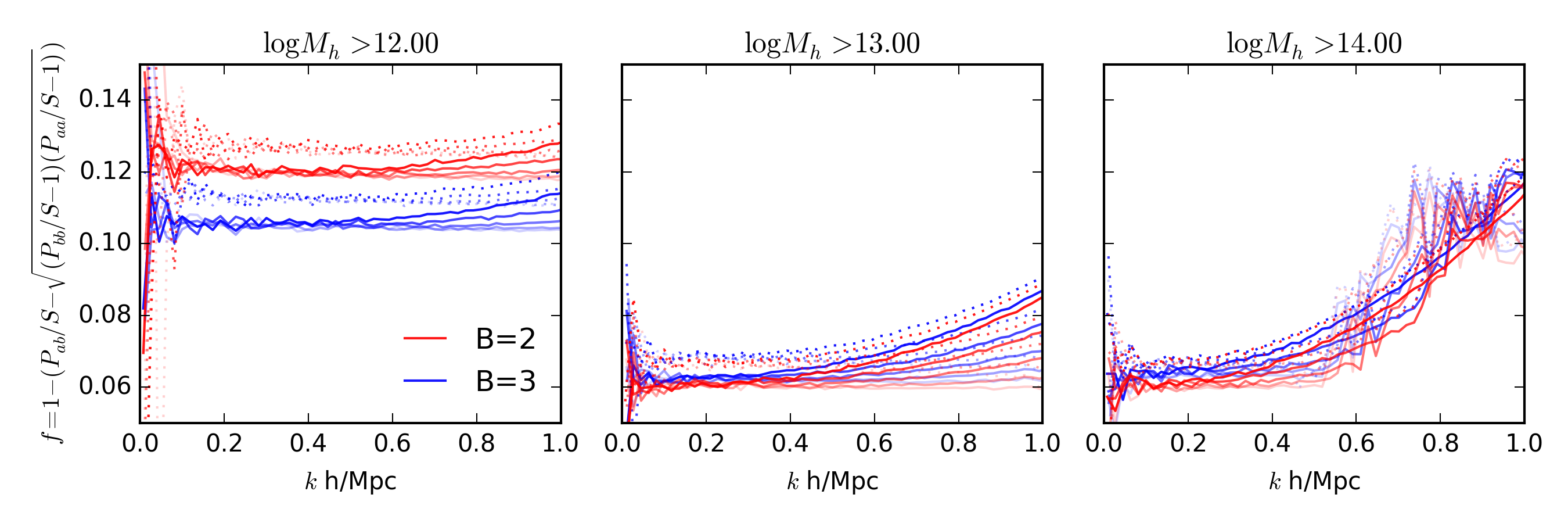}
\caption{Benchmarks on halos, varying force resolution. Top panel: transfer function. Center panel: cross correlation coefficient. Bottom panel: stochasticity. 
Red: $B=2$ ; Blue: $B=3$; Solid: PM; Dots: COLA. Opacity: Line of sight direction, $\mu$ = 0.1 (transparent), 0.3, 0.5, 0.7, 0.9 (opaque). Solid : PM. Dashed : COLA. $S=1/n$ is the shot-noise level.}
\label{fig:benchmarks-10}
\end{figure*}

The rest of the benchmarks are calculated after applying abundance matching to reassign halo masses in the FastPM simulations. We show the rest of the benchmark suite in Figure \ref{fig:benchmarks-10}, which have been calculated after abundance matching. The structure of the figure is similar to Figure \ref{fig:benchmarks}, but now we vary the force resolution. We again observe some slight advantages of PM comparing to COLA (one percent level). The improvement due to increasing the force resolution from $B=2$ to $B=3$ is very limited, typically at 1\% level. The $B=2$ approximation has a slightly larger (by 2\%) stochasticity than $B=3$ approximation for the least massive threshold ($M > 10^{12} h^{-1} M_\odot$). Given that $B=2$ simulation is almost 3 times faster than a $B=3$ simulation (Figure \ref{fig:runtime}), the 2\% increase in stochasticity is a reasonable price to pay.
Overall, we find that the $N_s=10$/$B=2$ approximation uses about 10\% of CPU time of a $N_s=40$/$B=3$ simulation, yet the benchmarks on halos of both are almost identical. 

\section{Conclusions}

In this paper we introduce FastPM, a new implementation of an approximate particle mesh N-body solver. FastPM modifies the standard kick and drift factors such that it agrees with Zeldovich solution on large scales, guaranteeing zero error in $k \rightarrow 0$ limit even for very few time steps. These modified factors assume acceleration during the time step is consistent with 1LPT, and they reduce to the standard ones in the limit of short time steps. We recommend the use of these modified factors whenever correct large scale evolution is desired. 

The domain decomposition in FastPM for parallel Fourier Transform and particle data is in 2-dimensions, allowing the code to scale almost linearly with the number of CPUs when a large number of CPUs (more than 10,000) are employed.

We then proceed to investigate numerical precision of halos identified with FoF within FastPM. Four benchmarks are defined, measuring the performance: ratio of halo mass function, transfer function, cross correlation coefficient, and stochasticity. We show that 
\begin{itemize} 
\item our implementation of PM (with modified kick and drift factors) performs slightly better than COLA on all benchmarks; especially when the number of time steps is low (5 steps).
\item the benchmarks on halos of any scheme with $N_s \ge 10$/$B \ge 2$ is very close to the exact solution.  This makes the $N_s = 10$/$B=2$ approximation very interesting, as it uses only 7 times of the computing time of generating a 2LPT initial condition, and 10\% of time of a $N_s = 40$/$B =3 $ approximated run. For higher redshifts, or for cases where large stochasticity can be tolerated, the $N_s = 5$/$B=2$ 
can also be adequate, at only 4 times the computing time of a 2LPT initial condition. 
\end{itemize}

We see several use cases for FastPM:
\begin{itemize}
\item As the halo catalog step of a mock factory, FastPM can be useful for generating a large number of mock catalogs. This is the same use case scenario similar to other recently proposed approximate N-body codes  \citep{2016MNRAS.459.2327I,lpicola2015,sunayama15}. 

\item Non-linear power spectrum (and higher order statistics) emulator: compared to other codes
FastPM can efficiently utilize a very large amount of computing resources. In fact, the turn around time with FastPM can be as low as 1 minute if sufficient computing resources are reserved for real time usage. This means a large number of models, varying both cosmological parameters and nuisance parameters (such as halo occupation distribution parameters), can be explored rapidly. 
We plan to implement and expose a programming interface for various cosmology models in FastPM.

\item Initial conditions solver: FastPM is designed as a software library, making it easy to embed into another application. One option is the 
initial conditions solver (\cite{wang2013}), which requires derivatives as a function of initial modes. For the derivatives to be computationally feasible one needs a simple force evaluation scheme and very few time steps, both satisfied by FastPM. The scheme we have proposed, $B=2$, $N_s=10$, uses 7 times more CPU time than a single 2LPT step, but provides realistic friend-of-friend halos. Depending of the allowed stochasticity budget, using $N_s=5$ or lower may also prove useful for computing the derivatives, given that the complexity of derivatives scales with $N_s$. 

\item While we have focused on halos in this paper, FastPM can also be used for other applications, such as weak lensing (where dark matter 
is used). When it comes to dark matter our recommended strategy is to multiply the density field with the transfer function, which can be obtained from FastPM itself ran at a higher resolution and number of time steps, or from a higher resolution simulation. Since the cross-correlation coefficient is very close to 1 up to $k=1{\rm h/Mpc}$, this would therefore guarantee high precision at least up to that $k$ value. We expect the transfer function to be a slowly varying function of cosmological parameters and redshift. One possible strategy is to calibrate FastPM against at sufficient number of points in parameter space to make this error negligible.  Finally, we expect the loss of precision at even higher k to be related to the structure of halos at small scales: FastPM already finds all the halos with the correct central position, and their mass error is relatively small, so only the internal halo profiles are incorrect. Since these are affected by baryonic effects (gas profile, AGN feedback etc.) anyways this needs to be addressed in any pure dark matter code, and low resolution FastPM is not necessarily limited relative to a high resolution N-body code. 
Another potential application is Lyman alpha forest, where a nonlinear transformation of matter density can be used as an approximation to the 
Lyman-alpha flux,
with all three applications above being of possible interest. We plan to present some of these applications in the future.
\end{itemize}

{\bf Acknowledgment}

We acknowledge support of NASA grant NNX15AL17G. 
The majority of the computing resources are provided at NERSC through the allocations for the Baryon Oscillation Spectroscopic Survey (BOSS) program and for the Berkeley Institute for Data Science (BIDS) program.
We thank Dr. Jun Kuda for distributing the source code of \url{cola_halo} under the GPLv3 license, which served both as a design inspiration and as a reference implementation of the COLA scheme.
We thank Alejandro Cervantes and Marcel Schmittfull for their generous help in testing the code. 
We thank Martin White for providing the RunPB TreePM simulation and initial condition that formed the foundations of our benchmarks.
The development of FastPM is hosted by \url{github.com} at \url{https://github.com/rainwoodman/fastpm}. 
The version of code used in this paper is based on commit \url{ 9219d0}. 
The data analysis software \url{nbodykit} is used for identifying Friend-of-Friend halos and calculating of power spectra. \footnote{\url{https://github.com/bccp/nbodykit}, separate publication being prepared.} We welcome collaboration on development of both software packages.

\appendix
\section{Choice of Differentiation Scheme}
\label{app:kernel}
\begin{figure}
\includegraphics[width=\columnwidth]{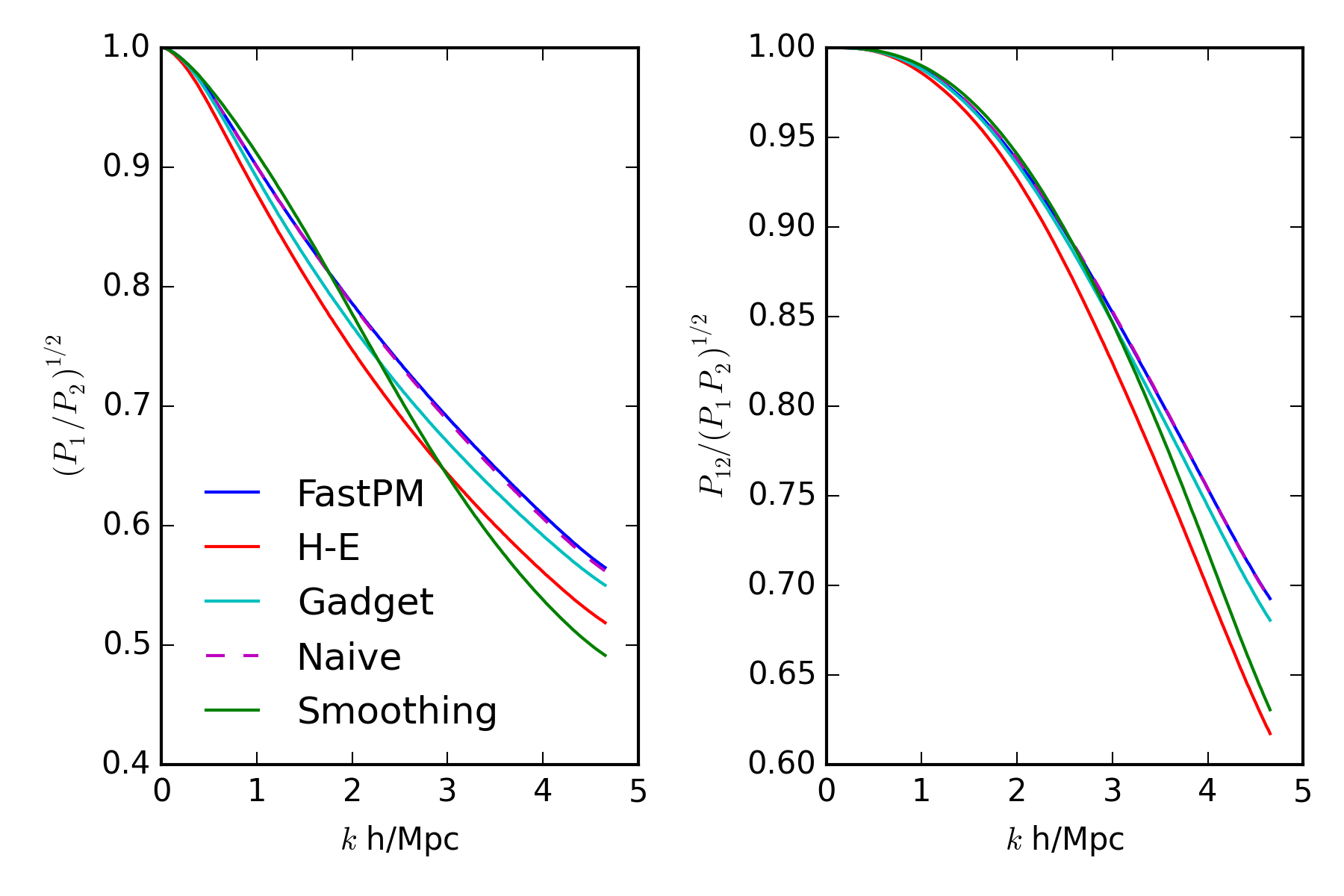}
\includegraphics[width=\columnwidth]{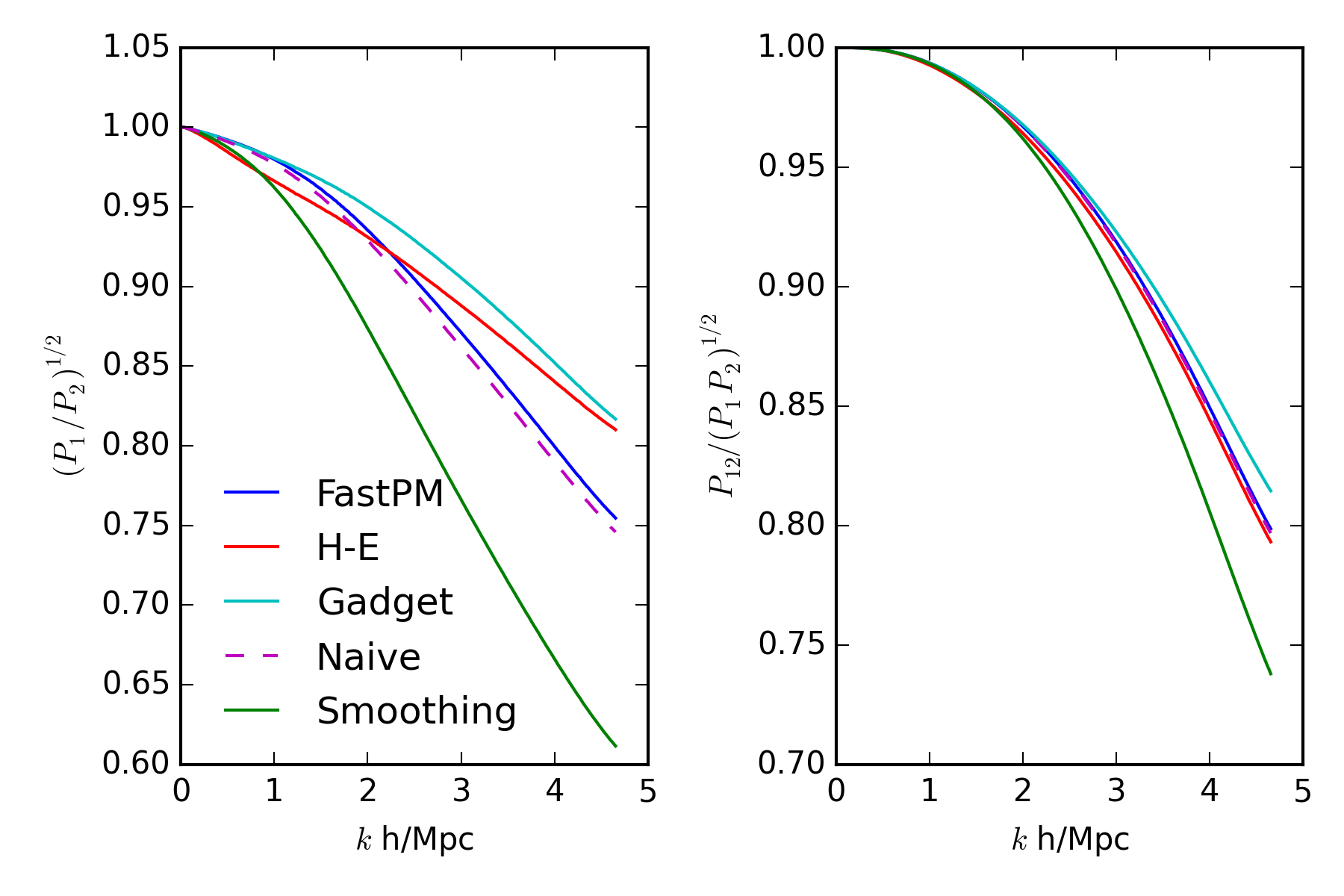}
\caption{Comparing $z=0$ power spectrum resulted from different schemes on the same initial condition with 10 and 40 step simulations. Upper panels : 10 steps; Lower panels: 40 steps. The reference is the RunPB simulation. The configuration of the simulations is the same as in Section 3, starting from $z=9$.}
\label{fig:kernels}
\end{figure}

\begin{figure}
\includegraphics[width=\columnwidth]{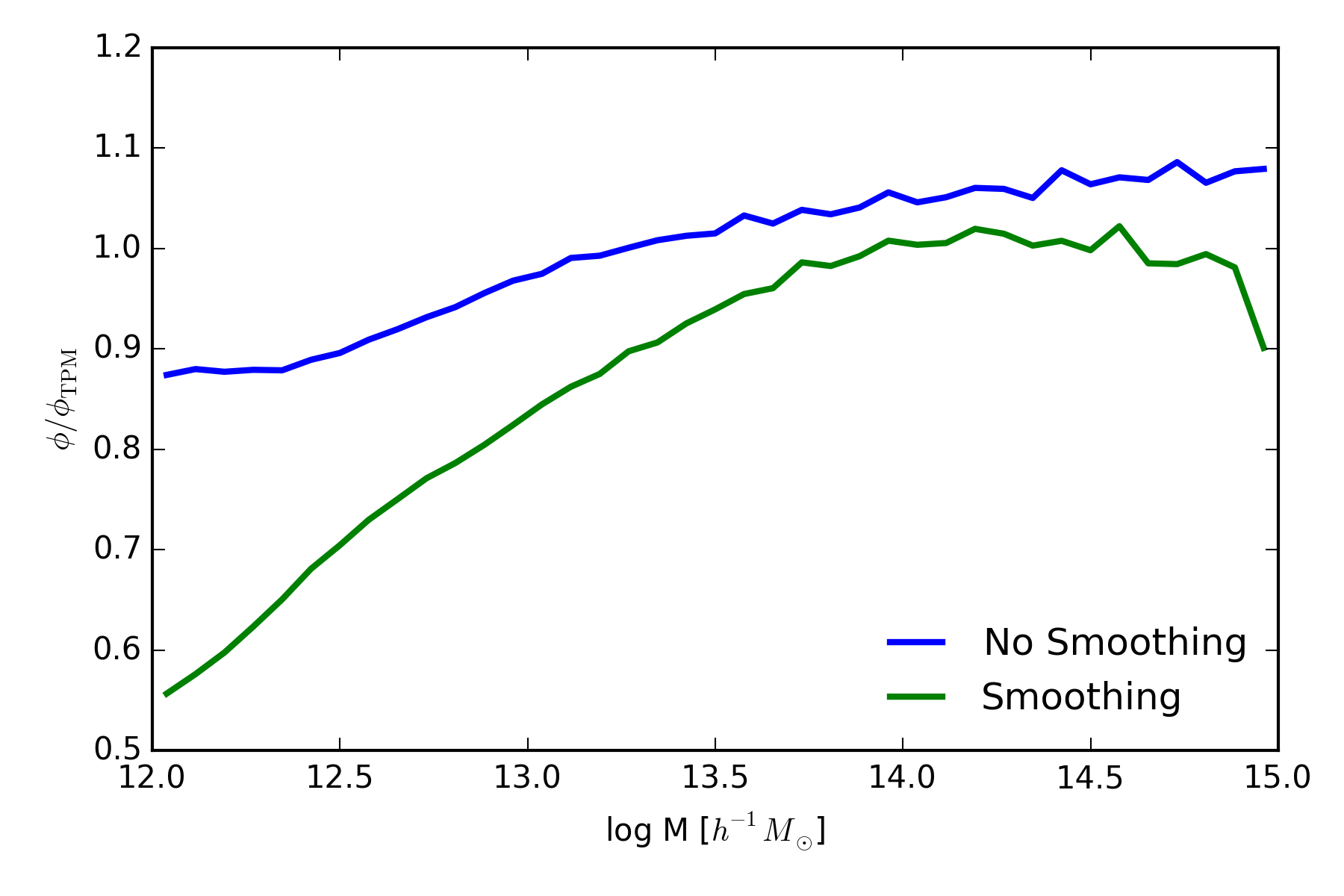}
\caption{Comparing $z=0$ mass function with and without smoothing on the same initial condition with 10 step simulations. The reference is the RunPB simulation. Blue: FastPM scheme without smoothing; Green: With smoothing. The configuration of the simulations is the same as Section 3.} 
\label{fig:kernels-mf}
\end{figure}

FastPM supports a variety of differentiation schemes. (Section 2). We explore the effect on dark matter density field due to choice of scheme in this section. This effect is shown in Figure \ref{fig:kernels}, where we compare the $z=0$ power spectrum resulted from different schemes on the same initial condition with 10 step and 40 step simulations against the RunPB simulation. The configuration of the simulations is the same as in Section 3, with a force mesh resolution of $L/ N_m = 0.34 h^{-1}\rm{Mpc}$. 
Various schemes show overall a good level of agreement at $k < 0.5 h/\rm{Mpc}$. The difference increases to 5 percent at $k = 2 h/\rm{Mpc}$. At 10 steps, the FastPM kernel and the Naive kernel produces the largest power and the highest cross correction coefficient. At 40 steps, the Gadget kernel shows improvements over FastPM and Naive kernel at $k > 1 h/\rm{Mpc}$.
Smoothing (Gaussian with $r_\mathrm{s} = L / N_m$) slightly increases the cross correlation coefficient at the cost of severely suppressing the power on small scales that corresponds to the collapse of halos, losing halos below $M < 10^{13.5} h^{-1}M_\odot$ (See Figure \ref{fig:kernels-mf}).

\section{Choice of Starting redshift}
\begin{figure}
\includegraphics[width=\columnwidth]{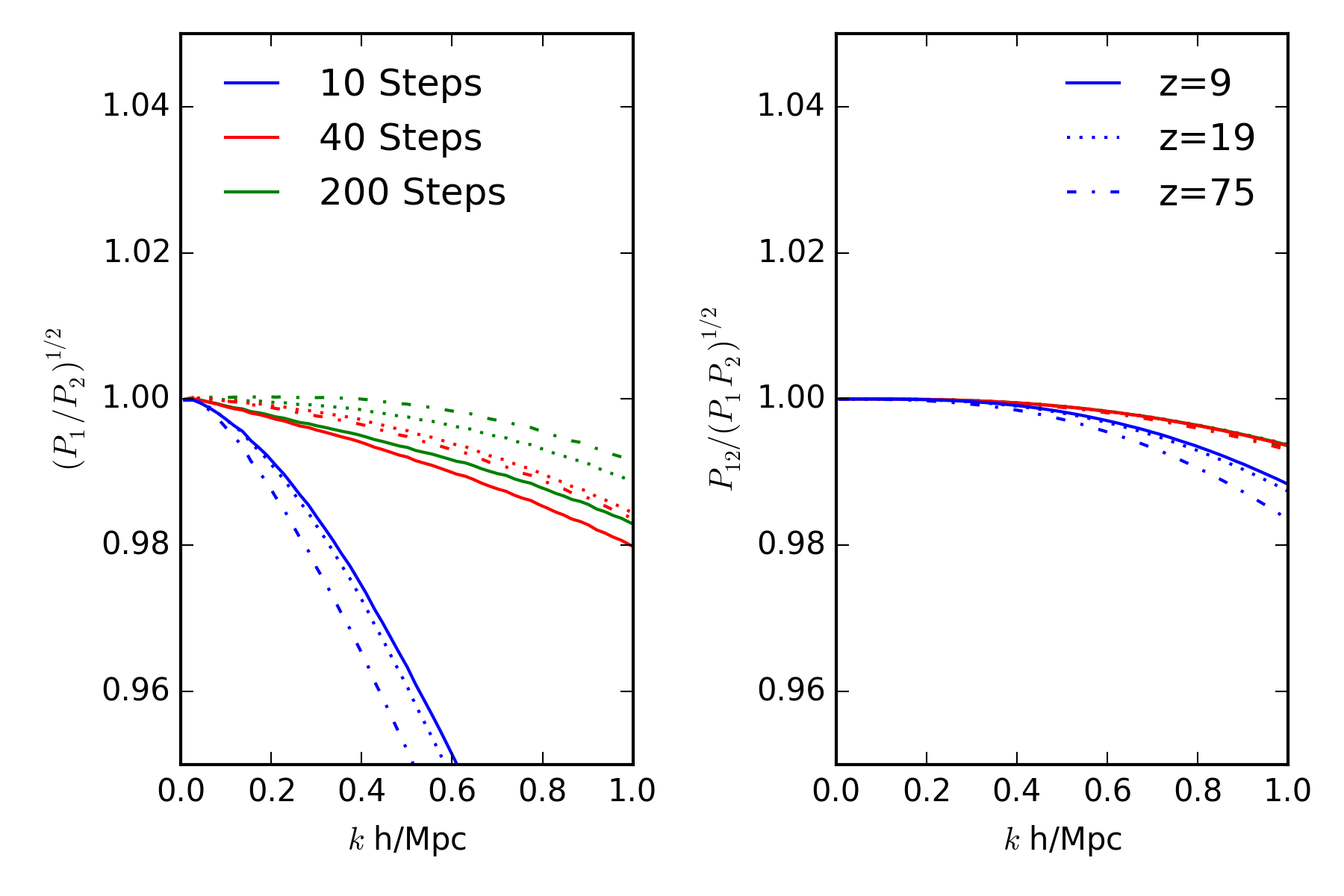}
\caption{Comparing $z=0$ power spectrum resulted from different starting redshifts on the same initial condition with 10 (blue), 40 (red), and 200 (green) step simulations. We show three starting redshifts, $z_i=$9 (solid), 19 (dotted), 75 (dotted-dashed). The reference is the RunPB simulation, which started at redshift $z_i=75$. Left: the square root of the ratio of power spectrum. Right: the cross correlation coefficient. Note that in the right panel the 200 step lines are covered by the 40 step lines.}
\label{fig:startingz}
\end{figure}
Starting redshift affects the accuracy of simulations by two competing effects. 

1) The temporal resolution decreases as the starting redshift increases. This effect is the most evident in the 10 step simulations, where the time steps are coarse. We see that the $z_i=75$ simulation performed worse than $z_i=9$ and $z_i=19$. Therefore, for simulations with very few time steps, a lower starting redshift improves the benchmarks.

2) The stochasticity in initial condition decreases as the starting redshift increases. In our case we want to minimize it against that of the reference simulation. We see this with the 200 steps simulation, where the $z_i=75$ 
simulation, which is the starting redshift of reference TreePM simulation, has the least error in power spectrum. The improvement due to better agreement in initial condition only affects the power spectrum, not the cross-correlation coefficient.
Therefore, for simulations with many time steps, matching the starting redshift to that of the reference N-body simulation gives the best benchmarks. This is not surprising because the large scale gravity force in the reference N-body simulation is calculated with the same Particle Mesh method, although with many more time steps. The effect between $z_i=19$ and $z_i=75$ is about 0.4\% in the power spectrum at $k=1{\rm h/Mpc}$ at $z=0$. 

\section{Choice of Linking Length}
\label{app:linking}
\begin{figure*}
\includegraphics[width=\textwidth]{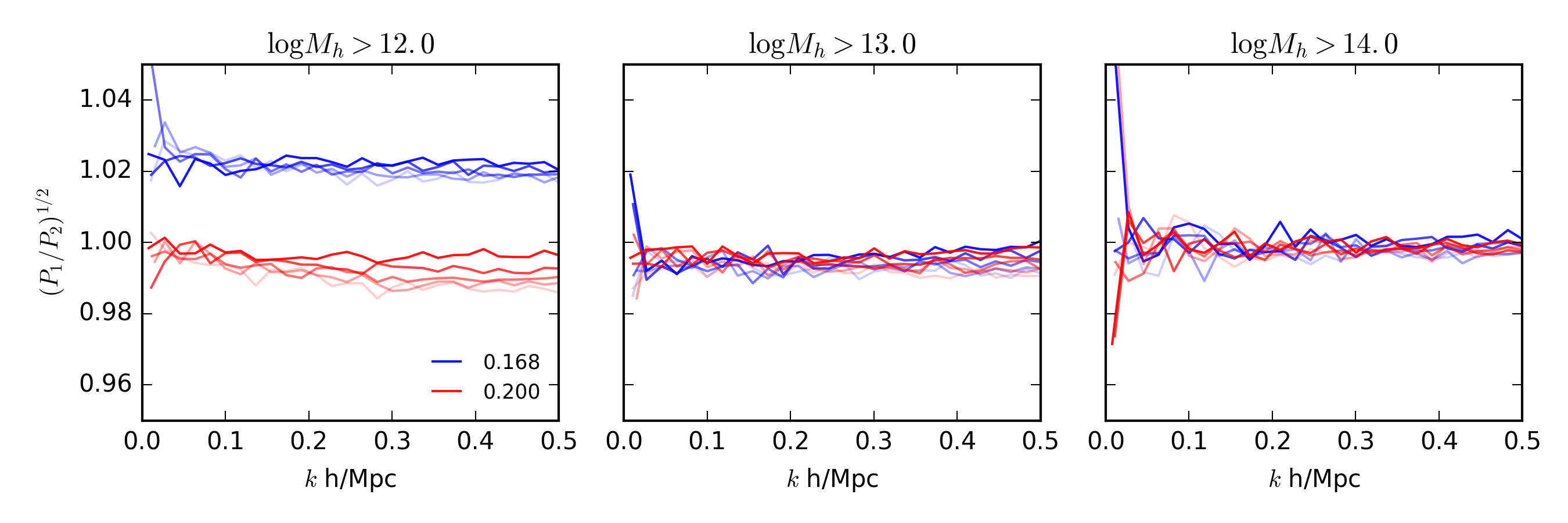}
\includegraphics[width=\textwidth]{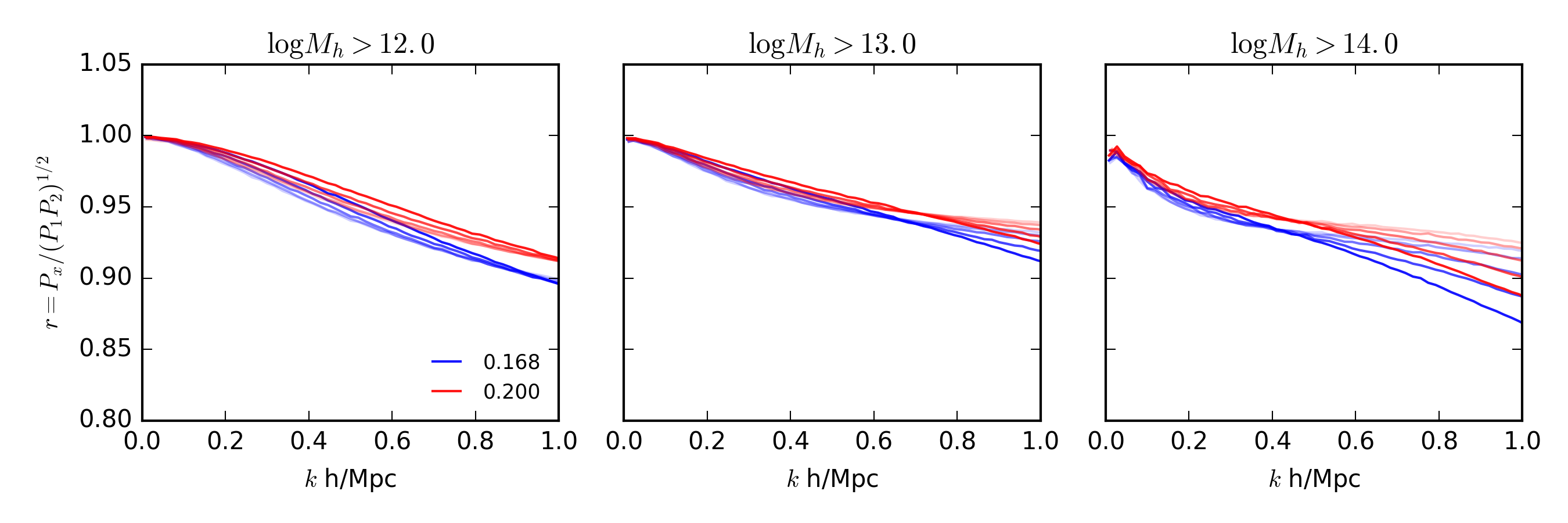}
\includegraphics[width=\textwidth]{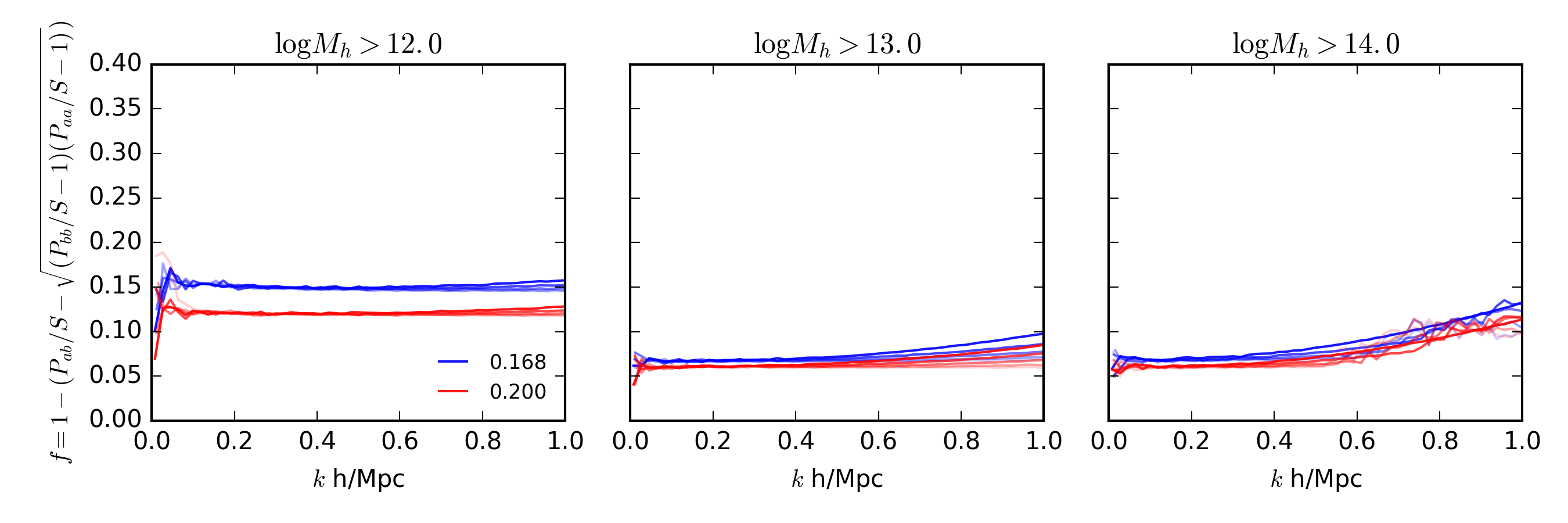}
\caption{Benchmarks on halos, varying linking length. Top panel: transfer function. Center panel: cross correlation coefficient. Bottom panel: stochasticity. 
Red: 0.200 ; Blue: 0.168; Opacity: Line of sight direction, $mu$ = 0.1 (transparent), 0.3, 0.5, 0.7, 0.9 (opaque). $S=1/n$ is the shot-noise level.}
\label{fig:ll0168}
\end{figure*}

The friends-of-friend algorithm that we use to identify objects are known to bridge halos into the same object. The effect can affect our halo detection because in FastPM the density field has less contrast than an accurate N-body simulation. Reducing linking length to 0.168 can alleviate this problem and improve the quality of the mock halo catalog \citep{2008ApJ...688..709T}. In Figure \ref{fig:ll0168} we show that reducing linking length from 0.2 to 0.168 does not affect the benchmarks, except for the least massive bin ($10^{12} h^{-1} M_\odot$), where we observe a constant 2\% increase in bias and stochasticity.

\bibliographystyle{mn2e}
\bibliography{mybib}
\end{document}